\pdfoutput=1
\documentclass[12pt,a4paper]{article}
\usepackage[graphicx]{realboxes}

\usepackage{ifthen} % for conditional statements
\newboolean{pdflatex}
\setboolean{pdflatex}{true} % False for eps figures 

\newboolean{articletitles}
\setboolean{articletitles}{true} % False removes titles in references

\newboolean{uprightparticles}
\setboolean{uprightparticles}{false} %True for upright particle symbols

\def\paperauthors{LHCb collaboration} % Leave as is for PAPER and CONF
\def\paperasciititle{Measurement of CP violation in the Bs->phi phi decay and search for the Bd->phi phi decay}
\def\papertitle{Measurement of \CP violation in the \BsPP decay and search for the $\Bd\to\phi\phi$ decay}
\def\paperkeywords{{High Energy Physics}, {LHCb}} % Comma separated list
\def\papercopyright{\the\year\ CERN for the benefit of the LHCb collaboration} % new since 9/Apr/2018
\def\paperlicence{CC-BY-4.0 licence}
\def\paperlicenceurl{https://creativecommons.org/licenses/by/4.0/}

\usepackage[top=1in, bottom=1.25in, left=1in, right=1in]{geometry}

\usepackage{dcolumn}
\newcolumntype{d}[1]{D{.}{.}{#1} }

\usepackage{microtype}
\usepackage{amsbsy}
\usepackage{verbatim}
\usepackage{rotating}
\usepackage{multirow}
\usepackage{lineno}  % for line numbering during review
\usepackage{xspace} % To avoid problems with missing or double spaces after
\usepackage{caption} %these three command get the figure and table captions automatically small

\usepackage{graphicx}  % to include figures (can also use other packages)
\usepackage{color}
\usepackage{colortbl}
\graphicspath{{./figs/}} % Make Latex search fig subdir for figures
\DeclareGraphicsExtensions{.pdf,.PDF,png,.PNG}

\usepackage{amsmath} % Adds a large collection of math symbols
\usepackage{amssymb}
\usepackage{amsfonts}
\usepackage{upgreek} % Adds in support for greek letters in roman typeset

\newcommand*\patchAmsMathEnvironmentForLineno[1]{%
\expandafter\let\csname old#1\expandafter\endcsname\csname #1\endcsname
\expandafter\let\csname oldend#1\expandafter\endcsname\csname
end#1\endcsname
 \renewenvironment{#1}%
   {\linenomath\csname old#1\endcsname}%
   {\csname oldend#1\endcsname\endlinenomath}%
}
\newcommand*\patchBothAmsMathEnvironmentsForLineno[1]{%
  \patchAmsMathEnvironmentForLineno{#1}%
  \patchAmsMathEnvironmentForLineno{#1*}%
}
\AtBeginDocument{%
\patchBothAmsMathEnvironmentsForLineno{equation}%
\patchBothAmsMathEnvironmentsForLineno{align}%
\patchBothAmsMathEnvironmentsForLineno{flalign}%
\patchBothAmsMathEnvironmentsForLineno{alignat}%
\patchBothAmsMathEnvironmentsForLineno{gather}%
\patchBothAmsMathEnvironmentsForLineno{multline}%
\patchBothAmsMathEnvironmentsForLineno{eqnarray}%
}

\usepackage{hyperxmp}

\usepackage[pdftex,
            pdfauthor={\paperauthors},
            pdftitle={\paperasciititle},
            pdfkeywords={\paperkeywords},
            pdfcopyright={Copyright (C) \papercopyright},
            pdflicenseurl={\paperlicenceurl}]{hyperref}

\usepackage[colorinlistoftodos,textsize=scriptsize]{todonotes}

\usepackage[all]{hypcap} % Internal hyperlinks to floats.

\usepackage{xspace} 
\usepackage{upgreek}

\def\lhcb   {\mbox{LHCb}\xspace}

\def\cdf    {\mbox{CDF}\xspace}

\def\bfactories {\mbox{\B Factories}\xspace}

\def\MagUp {\mbox{\em Mag\kern -0.05em Up}\xspace}

\ifthenelse{\boolean{uprightparticles}}%
{

 \def\Pmu         {\ensuremath{\upmu}\xspace}                 
 \def\Pnu         {\ensuremath{\upnu}\xspace}                 
                  
 \def\Ppi         {\ensuremath{\uppi}\xspace}

 \def\Pphi        {\ensuremath{\upphi}\xspace}

 \def\Ppsi        {\ensuremath{\uppsi}\xspace}

 \def\PDelta      {\ensuremath{\Delta}\xspace}                 
 \def\PXi         {\ensuremath{\Xi}\xspace}                 
 \def\PLambda     {\ensuremath{\Lambda}\xspace}                 
 \def\PSigma      {\ensuremath{\Sigma}\xspace}                 
 \def\POmega      {\ensuremath{\Omega}\xspace}                 
 \def\PUpsilon    {\ensuremath{\Upsilon}\xspace}

 \def\PB      {\ensuremath{\mathrm{B}}\xspace}                 
                  
 \def\PD      {\ensuremath{\mathrm{D}}\xspace}

 \def\PJ      {\ensuremath{\mathrm{J}}\xspace}                 
 \def\PK      {\ensuremath{\mathrm{K}}\xspace}

 \def\Pb      {\ensuremath{\mathrm{b}}\xspace}                 
 \def\Pc      {\ensuremath{\mathrm{c}}\xspace}

 \def\Pi      {\ensuremath{\mathrm{i}}\xspace}

 \def\Pp      {\ensuremath{\mathrm{p}}\xspace}

 \def\Ps      {\ensuremath{\mathrm{s}}\xspace}

 \def\thebaroffset{0.0em}
}
{

 \def\Pmu         {\ensuremath{\mu}\xspace}                 
 \def\Pnu         {\ensuremath{\nu}\xspace}                 
                  
 \def\Ppi         {\ensuremath{\pi}\xspace}

 \def\Pphi        {\ensuremath{\phi}\xspace}

 \def\Ppsi        {\ensuremath{\psi}\xspace}                 
                  
 \mathchardef\PDelta="7101
 \mathchardef\PXi="7104
 \mathchardef\PLambda="7103
 \mathchardef\PSigma="7106
 \mathchardef\POmega="710A
 \mathchardef\PUpsilon="7107
                  
 \def\PB      {\ensuremath{B}\xspace}                 
                  
 \def\PD      {\ensuremath{D}\xspace}

 \def\PJ      {\ensuremath{J}\xspace}                 
 \def\PK      {\ensuremath{K}\xspace}

 \def\Pb      {\ensuremath{b}\xspace}                 
 \def\Pc      {\ensuremath{c}\xspace}

 \def\Pi      {\ensuremath{i}\xspace}

 \def\Pp      {\ensuremath{p}\xspace}

 \def\Ps      {\ensuremath{s}\xspace}

 \def\thebaroffset{0.18em}
}
\newcommand{\offsetoverline}[2][\thebaroffset]{\kern #1\overline{\kern -#1 #2}}%

\makeatletter
\ifcase \@ptsize \relax% 10pt
  \newcommand{\miniscule}{\@setfontsize\miniscule{4}{5}}% \tiny: 5/6
\or% 11pt
  \newcommand{\miniscule}{\@setfontsize\miniscule{5}{6}}% \tiny: 6/7
\or% 12pt
  \newcommand{\miniscule}{\@setfontsize\miniscule{5}{6}}% \tiny: 6/7
\fi
\makeatother

\DeclareRobustCommand{\optbar}[1]{\shortstack{{\miniscule (\rule[.5ex]{1.25em}{.18mm})}
  \\ [-.7ex] $#1$}}

\def\mup        {{\ensuremath{\Pmu^+}}\xspace}
\def\mun        {{\ensuremath{\Pmu^-}}\xspace} % muon negative (\mum is taken)

\def\neu        {{\ensuremath{\Pnu}}\xspace}

\def\neum       {{\ensuremath{\neu_\mu}}\xspace}

\def\squark    {{\ensuremath{\Ps}}\xspace}
\def\squarkbar {{\ensuremath{\overline \squark}}\xspace}
\def\ssbar     {{\ensuremath{\squark\squarkbar}}\xspace}
\def\cquark    {{\ensuremath{\Pc}}\xspace}
\def\cquarkbar {{\ensuremath{\overline \cquark}}\xspace}
\def\ccbar     {{\ensuremath{\cquark\cquarkbar}}\xspace}
\def\bquark    {{\ensuremath{\Pb}}\xspace}
\def\bquarkbar {{\ensuremath{\overline \bquark}}\xspace}

\def\pion   {{\ensuremath{\Ppi}}\xspace}

\def\pip    {{\ensuremath{\pion^+}}\xspace}
\def\pim    {{\ensuremath{\pion^-}}\xspace}

\def\kaon    {{\ensuremath{\PK}}\xspace}
\def\Kbar    {{\ensuremath{\offsetoverline{\PK}}}\xspace}

\def\KorKbar {\kern \thebaroffset\optbar{\kern -\thebaroffset \PK}{}\xspace}

\def\Kp      {{\ensuremath{\kaon^+}}\xspace}
\def\Km      {{\ensuremath{\kaon^-}}\xspace}

\def\Kstarz  {{\ensuremath{\kaon^{*0}}}\xspace}
\def\Kstarzb {{\ensuremath{\Kbar{}^{*0}}}\xspace}
\def\Kstar   {{\ensuremath{\kaon^*}}\xspace}

\newcommand{\phiz}{\ensuremath{\Pphi}\xspace}

\def\Dbar    {{\ensuremath{\offsetoverline{\PD}}}\xspace}
\def\D       {{\ensuremath{\PD}}\xspace}

\def\DorDbar {\kern \thebaroffset\optbar{\kern -\thebaroffset \PD}\xspace}

\def\Dzb     {{\ensuremath{\Dbar{}^0}}\xspace}

\def\Dstarm  {{\ensuremath{\D^{*-}}}\xspace}

\def\Dsm     {{\ensuremath{\D^-_\squark}}\xspace}

\def\B       {{\ensuremath{\PB}}\xspace}
\def\Bbar    {{\ensuremath{\offsetoverline{\PB}}}\xspace}

\def\BorBbar {\kern \thebaroffset\optbar{\kern -\thebaroffset \PB}\xspace}
\def\Bz      {{\ensuremath{\B^0}}\xspace}

\def\Bd      {{\ensuremath{\B^0}}\xspace}

\def\BdorBdbar {\kern \thebaroffset\optbar{\kern -\thebaroffset \Bd}\xspace}
\def\Bu      {{\ensuremath{\B^+}}\xspace}

\def\Bp      {{\ensuremath{\Bu}}\xspace}

\def\Bs      {{\ensuremath{\B^0_\squark}}\xspace}
\def\Bsb     {{\ensuremath{\Bbar{}^0_\squark}}\xspace}
\def\BsorBsbar {\kern \thebaroffset\optbar{\kern -\thebaroffset \Bs}\xspace}

\def\jpsi     {{\ensuremath{{\PJ\mskip -3mu/\mskip -2mu\Ppsi\mskip 2mu}}}\xspace}

\def\Y#1S{\ensuremath{\PUpsilon{(#1S)}}\xspace}

\def\proton      {{\ensuremath{\Pp}}\xspace}

\def\Lz          {{\ensuremath{\PLambda}}\xspace}

\def\LorLbar     {\kern \thebaroffset\optbar{\kern -\thebaroffset \PLambda}\xspace}

\def\Lb           {{\ensuremath{\Lz^0_\bquark}}\xspace}

\newcommand{\decay}[2]{\ensuremath{#1\!\to #2}\xspace}         % Not mbox! It changes sizes

\def\to                 {\ensuremath{\rightarrow}\xspace}

\def\CP                {{\ensuremath{C\!P}}\xspace}

\def\T                 {{\ensuremath{T}}\xspace}

\newcommand{\dms}{{\ensuremath{\Delta m_{\squark}}}\xspace}

\newcommand{\DGs}{{\ensuremath{\Delta\Gamma_{\squark}}}\xspace}

\newcommand{\Gs}{{\ensuremath{\Gamma_{\squark}}}\xspace}

\def\AT#1     {\ensuremath{A_{\mathrm{T}}^{#1}}\xspace}           % 2

\def\C#1      {\ensuremath{\mathcal{C}_{#1}}\xspace}                       % 9
\def\Cp#1     {\ensuremath{\mathcal{C}_{#1}^{'}}\xspace}                    % 7
\def\Ceff#1   {\ensuremath{\mathcal{C}_{#1}^{\mathrm{(eff)}}}\xspace}        % 9  
\def\Cpeff#1  {\ensuremath{\mathcal{C}_{#1}^{'\mathrm{(eff)}}}\xspace}       % 7
\def\Ope#1    {\ensuremath{\mathcal{O}_{#1}}\xspace}                       % 2
\def\Opep#1   {\ensuremath{\mathcal{O}_{#1}^{'}}\xspace}                    % 7

\newcommand{\nospaceunit}[1]{\ensuremath{\text{#1}}}       
\newcommand{\aunit}[1]{\ensuremath{\text{\,#1}}}       
\newcommand{\tev}{\aunit{Te\kern -0.1em V}\xspace}
\newcommand{\gev}{\aunit{Ge\kern -0.1em V}\xspace}
\newcommand{\mev}{\aunit{Me\kern -0.1em V}\xspace}
\newcommand{\kev}{\aunit{ke\kern -0.1em V}\xspace}
\newcommand{\ev}{\aunit{e\kern -0.1em V}\xspace}
\newcommand{\mevc}{\ensuremath{\aunit{Me\kern -0.1em V\!/}c}\xspace}
\newcommand{\gevc}{\ensuremath{\aunit{Ge\kern -0.1em V\!/}c}\xspace}
\newcommand{\mevcc}{\ensuremath{\aunit{Me\kern -0.1em V\!/}c^2}\xspace}
\newcommand{\gevcc}{\ensuremath{\aunit{Ge\kern -0.1em V\!/}c^2}\xspace}
\def\mum  {\ensuremath{\,\upmu\nospaceunit{m}}\xspace}

\def\fb   {\ensuremath{\aunit{fb}}\xspace}
\def\invfb   {\ensuremath{\fb^{-1}}\xspace}

\def\ps   {\ensuremath{\aunit{ps}}\xspace}
\def\fs   {\aunit{fs}}

\def\invps{\ensuremath{\ps^{-1}}\xspace}

\newcommand{\stat}{\aunit{(stat)}\xspace}
\newcommand{\syst}{\aunit{(syst)}\xspace}

\newcommand{\chisq}{\ensuremath{\chi^2}\xspace}

\newcommand{\chisqip}{\ensuremath{\chi^2_{\text{IP}}}\xspace}

\def\deriv {\ensuremath{\mathrm{d}}}

\def\gsim{{~\raise.15em\hbox{$>$}\kern-.85em
          \lower.35em\hbox{$\sim$}~}\xspace}
\def\lsim{{~\raise.15em\hbox{$<$}\kern-.85em
          \lower.35em\hbox{$\sim$}~}\xspace}

\def\sPlot{\mbox{\em sPlot}\xspace}

\def\pt         {\ensuremath{p_{\mathrm{T}}}\xspace}

\def\ptot       {\ensuremath{p}\xspace}

\def\rad{\aunit{rad}}

\def\evtgen     {\mbox{\textsc{EvtGen}}\xspace}

\def\geant      {\mbox{\textsc{Geant4}}\xspace}

\def\photos     {\mbox{\textsc{Photos}}\xspace}

\def\pythia     {\mbox{\textsc{Pythia}}\xspace}

\xspace

\def\tell1  {TELL1\xspace}
\def\ukl1   {UKL1\xspace}

\newcommand{\ie}{\mbox{\itshape i.e.}\xspace}

\newcommand{\az}{\ensuremath{|A_0|^2}\xspace}

\newcommand{\dz}{\ensuremath{\delta_0}\xspace}
\newcommand{\ds}{\ensuremath{\delta_s}\xspace}
\newcommand{\dss}{\ensuremath{\delta_{ss}}\xspace}

\newcommand{\dpa}{\ensuremath{\delta_\parallel}\xspace}
\newcommand{\ape}{\ensuremath{|A_\perp|^2}\xspace}

\newcommand{\dpe}{\ensuremath{\delta_\perp}\xspace}
\newcommand{\philo}{\ensuremath{\phi_{s,0}}\xspace}
\newcommand{\phipa}{\ensuremath{\phi_{s,\parallel}}\xspace}
\newcommand{\phipe}{\ensuremath{\phi_{s,\perp}}\xspace}
\newcommand{\phisw}{\ensuremath{\phi_{s,s}}\xspace}
\newcommand{\phissw}{\ensuremath{\phi_{s,ss}}\xspace}
\newcommand{\lambz}{\ensuremath{|\lambda_{0}|}\xspace}
\newcommand{\lambpa}{\ensuremath{|\lambda_{\parallel}|}\xspace}
\newcommand{\lambpe}{\ensuremath{|\lambda_{\perp}|}\xspace}
\newcommand{\lambs}{\ensuremath{|\lambda_{s}|}\xspace}
\newcommand{\lambss}{\ensuremath{|\lambda_{ss}|}\xspace}
\def\BsPP         {\decay{\Bs}{\phi\phi}}
\newcommand{\phisPP}{{\ensuremath{\phi_{\squark}^{\ssbar\squark}}}\xspace}
\usepackage{cite} % Allows for ranges in citations
\usepackage{mciteplus}

\usepackage{longtable} % only for template; not usually to be used in PAPERs

\begin{document}

%%%%%%%%%%%%%%%%%%%%%%%%%
%%%%% Title     %%%%%%%%%
%%%%%%%%%%%%%%%%%%%%%%%%%
\renewcommand{\thefootnote}{\fnsymbol{footnote}}
\setcounter{footnote}{1}

% %%%%%%% CHOOSE TITLE PAGE--------
%\onecolumn
%\input{title-LHCb-INT}
%\input{title-LHCb-ANA}
%\input{title-LHCb-CONF}
% $Id: title-LHCb-PAPER.tex 122889 2018-08-17 17:59:55Z pkoppenb $
% ===============================================================================
% Purpose: LHCb-PAPER journal paper title page template
% Author: 
% Created on: 2010-09-25
% ===============================================================================

%%%%%%%%%%%%%%%%%%%%%%%%%
%%%%%  TITLE PAGE  %%%%%%
%%%%%%%%%%%%%%%%%%%%%%%%%
\begin{titlepage}
\pagenumbering{roman}

% Header ---------------------------------------------------
\vspace*{-1.5cm}
\centerline{\large EUROPEAN ORGANIZATION FOR NUCLEAR RESEARCH (CERN)}
\vspace*{1.5cm}
\noindent
\begin{tabular*}{\linewidth}{lc@{\extracolsep{\fill}}r@{\extracolsep{0pt}}}
\ifthenelse{\boolean{pdflatex}}% Logo format choice
{\vspace*{-1.5cm}\mbox{\!\!\!\includegraphics[width=.14\textwidth]{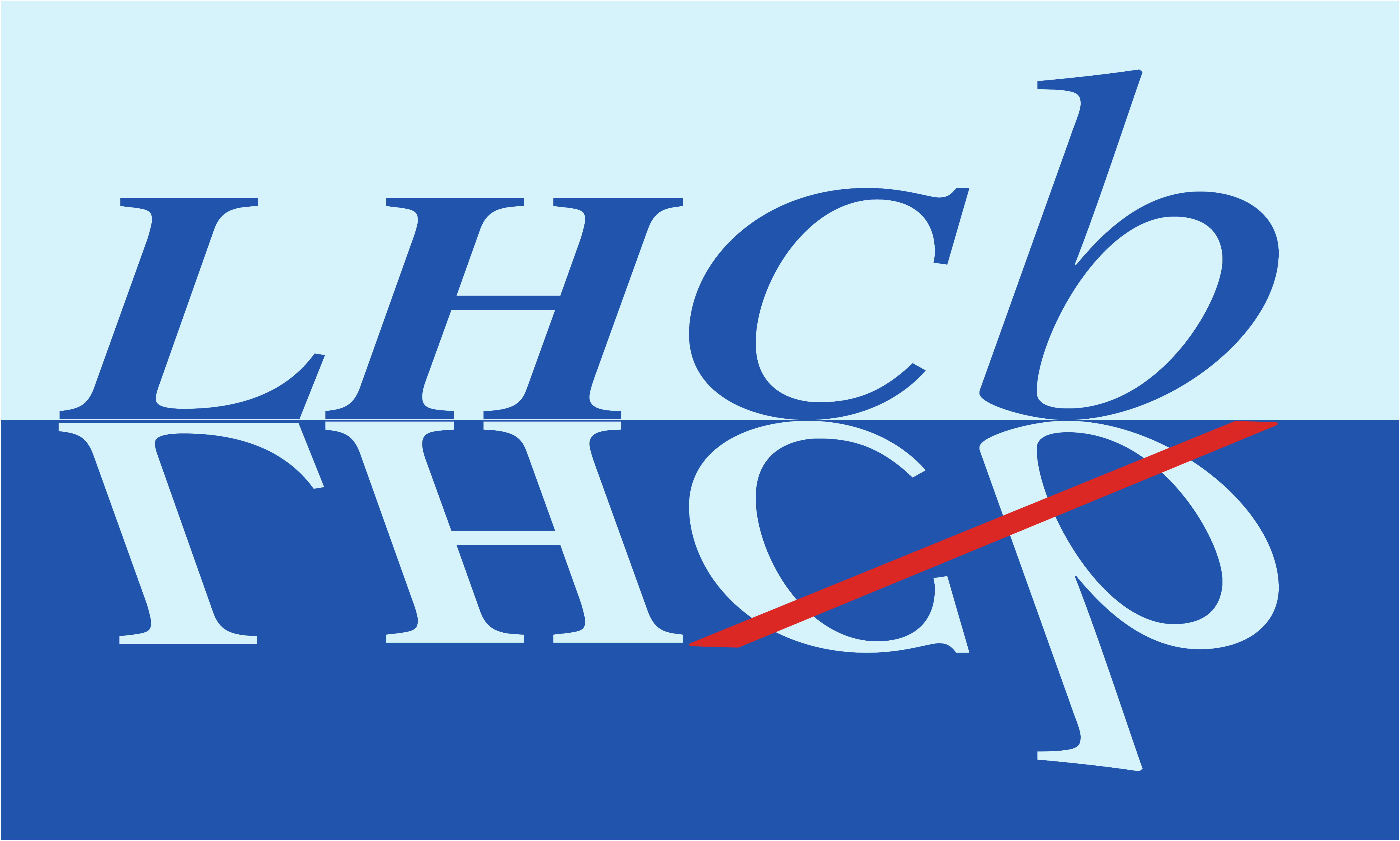}} & &}%
{\vspace*{-1.2cm}\mbox{\!\!\!\includegraphics[width=.12\textwidth]{figs/lhcb-logo.eps}} & &}%
\\
 & & CERN-EP-2019-121 \\  % ID 
 & & LHCb-PAPER-2019-019 \\  % ID 
 & & 23 July 2019 \\
 %& & \today \\ % Date - Can also hardwire e.g.: 23 March 2010
 & & \\
% not in paper \hline
\end{tabular*}

\vspace*{3.0cm}

% Title --------------------------------------------------
{\normalfont\bfseries\boldmath\huge
\begin{center}
% DO NOT EDIT HERE. Instead edit macro in main.tex to keep metadata correct
  \papertitle 
\end{center}
}

\vspace*{2.0cm}

% Authors -------------------------------------------------
\begin{center}
%In the footnote, replace 'paper' by 'Letter' in case of submission to PRL or PLB 
% Edit macro in main.tex to keep metadata correct
\paperauthors\footnote{Authors are listed at the end of this paper.}
\end{center}

\vspace{\fill}

% Abstract -----------------------------------------------
\begin{abstract}
  \noindent
  A measurement of the time-dependent \CP-violating asymmetry 
  in $\Bs \to \phi\phi$ decays is presented.
  %, along with measurements of the \T-odd
  %triple-product asymmetries. 
  Using a sample of proton-proton collision data
  corresponding to an integrated luminosity of $5.0\invfb$ collected by the \lhcb experiment at centre-of-mass energies $\sqrt{s} = 7$ \tev in 2011, 8 \tev in 2012 and 13 \tev in 2015 and 2016,
  a signal yield of around 9000 \BsPP decays is obtained. The
  % 8843
  \CP-violating phase $\phi_s^{s\bar{s}s}$ is measured to be $-0.073 \pm 0.115\stat\pm 0.027\syst\rad$, under the assumption it is independent of the helicity of the $\phi\phi$ decay.
  In addition, the
  \CP-violating phases of the transverse polarisations under the assumption of \CP conservation of the longitudinal phase are measured.
  The helicity-independent direct \CP-violation parameter is also measured, and is found to be $|\lambda|=0.99 \pm 0.05 \stat \pm 0.01 \syst$. In addition, $T$-odd triple-product
  asymmetries are measured. The results obtained are consistent with the
  hypothesis of \CP conservation in $\bquarkbar\to\squarkbar\ssbar$ transitions. Finally, a limit on the branching fraction of the $\Bd\to\phi\phi$ decay is determined to be $\mathcal{B}(\Bd\to\phi\phi)<2.7\times 10^{-8}$ at 90\,\% confidence level.
\end{abstract}

\vspace*{2.0cm}

\begin{center}
  Published in JHEP 12 (2019) 155
%  Phys.~Rev.~D /
%  Phys.~Rev.~Lett. /
%  Phys.~Lett.~B /
%  Eur.~Phys.~J.~C /
%  %  Nucl.~Phys.~B /
%  Chin.~Phys.~C /
%  Nature~Physics /
%  sciPost~Physics /
%  J. Instr. /
%  Instruments 
\end{center}

\vspace{\fill}

{\footnotesize 
% Edit macro in main.tex to keep metadata correct
\centerline{\copyright~\papercopyright. \href{\paperlicenceurl}{\paperlicence}.}}
\vspace*{2mm}

\end{titlepage}

%%%%%%%%%%%%%%%%%%%%%%%%%%%%%%%%
%%%%%  EOD OF TITLE PAGE  %%%%%%
%%%%%%%%%%%%%%%%%%%%%%%%%%%%%%%%

%  empty page follows the title page ----
\newpage
\setcounter{page}{2}
\mbox{~}
%\newpage
%
%% Author List ----------------------------
%%  You need to get a new author list!
%\input{LHCb_authorlist.tex}
%
%The author list for journal publications is provided by the Membership Committee shortly after 'approval to go to paper' has been given.
%%It will be made available on the page
%%\verb!http://www.physik.uzh.ch/~strauman/forMemCo/LHCb-PAPER-XXXX-XXX/! .
%It will be sent to you by email shortly after a paper number has beens assigned.
%The author list should be included already at first circulation, 
%to allow new members of the collaboration to verify whether they have been included correctly.
%Occasionally a misspelled name is corrected or associated institutions become full members.
%In that case, a new author list will be sent to you.
%In case line numbering doesn't work well after including the authorlist, try moving the \verb!\bigskip! after the last author to a separate line.
%
%
%The authorship for Conference Reports should be ``The LHCb
%  collaboration'', with a footnote giving the name(s) of the contact
%  author(s), but without the full list of collaboration names.

\cleardoublepage

%\twocolumn
% %%%%%%%%%%%%% ---------

\renewcommand{\thefootnote}{\arabic{footnote}}
\setcounter{footnote}{0}

%%%%%%%%%%%%%%%%%%%%%%%%%%%%%%%%
%%%%%  Table of Content   %%%%%%
%%%%%%%%%%%%%%%%%%%%%%%%%%%%%%%%
%%%% Uncomment next 2 lines if desired
%\tableofcontents
%\cleardoublepage

%%%%%%%%%%%%%%%%%%%%%%%%%
%%%%% Main text %%%%%%%%%
%%%%%%%%%%%%%%%%%%%%%%%%%

\pagestyle{plain} % restore page numbers for the main text
\setcounter{page}{1}
\pagenumbering{arabic}

%% Uncomment during review phase. 
%% Comment before a final submission.
%\linenumbers

% You can include short sections directly in the main tex file.
% However, for larger papers it is desirable to split the text into
% several semiautonomous files, which can be revised independently.
% This is especially useful when developing a document in
% collaboration with several people, since then different parts can be
% edited independently.  This type of file organization is shown here.
% 

\section{Introduction}
\label{sec:Introduction}

In the Standard Model (SM) the $\Bs\to\phi\phi$ decay, where the $\phi(1020)$ is implied throughout this paper, is forbidden at tree level and proceeds predominantly via a gluonic $\bquarkbar \to \squarkbar \ssbar$ loop (penguin) process. 
Hence, this channel provides an excellent probe of new heavy particles entering 
the penguin quantum loops~\cite{Bartsch:2008ps,Beneke:2006hg,PhysRevD.80.114026}.
In the SM, \CP violation is governed by a single phase in the Cabibbo--Kobayashi--Maskawa quark mixing matrix
\cite{Kobayashi:1973fv,*Cabibbo:1963yz}. 
Interference caused by the resulting weak phase difference between the \Bs-\Bsb oscillation and decay amplitudes leads to a \CP asymmetry
in the decay-time distributions of $\Bs$ and $\Bsb$ mesons.
%The \CP asymmetry is characterised by a \CP-violating weak phase.
%Due to different decay amplitudes, the actual value of the weak phase depends on the \Bs decay channel.
For  $\Bs \to \jpsi\Kp\Km$ and $\Bs \to \jpsi\pip\pim$ decays, which proceed via $\bquarkbar \to \squarkbar\ccbar$ transitions, the SM prediction of the weak phase
is $- 2 \arg \left( -V_{ts}^{ } V_{tb}^*/ V_{cs}^{ } V_{cb}^*\right)=-0.0369^{+0.0010}_{-0.0007}\; \rm rad$ according to the CKMfitter group~\cite{CKMfitter2015}, and $- 2 \arg \left( -V_{ts}^{ } V_{tb}^*/ V_{cs}^{ } V_{cb}^*\right)=-0.0370\pm 0.0010\; \rm rad$ according to the UTfit collaboration~\cite{UTfit-UT}.
The \lhcb collaboration has measured the weak phase in several decay processes: $\Bs \to \jpsi\Kp\Km$, $\Bs \to \jpsi\pip\pim$, $\Bs \to \jpsi\Kp\Km$ for the $\Kp\Km$ invariant mass region above 1.05 \gevc, $\Bs \to \psi(2S)\phi$ and $\Bs \to D_s^+ D_s^-$, corresponding to the combined result of $-0.041\pm 0.025\rad$~\cite{LHCb-PAPER-2019-013}. 
These measurements are consistent with the SM prediction and place stringent constraints on
\CP violation in \Bs-\Bsb oscillations~\cite{LHCb-PAPER-2012-031}. 
The \CP-violating phase, \phisPP, in the \BsPP decay is expected to be small in the SM.
Calculations using quantum chromodynamics factorisation (QCDf) 
provide an upper limit of $0.02\rad$ for its absolute value~\cite{Bartsch:2008ps,Beneke:2006hg,PhysRevD.80.114026}.
The previous most accurate measurement is $\phisPP=-0.17\pm0.15\stat\pm0.03\syst\rad$~\cite{LHCb-PAPER-2014-026}.

\CP violation can also be probed by time-integrated triple-product asymmetries.
These are formed from \T-odd combinations of the momenta of the final-state particles. These asymmetries complement the decay-time-dependent
measurement~\cite{gronau} and are expected to be close to zero in the SM~\cite{Datta:2012ky}. 
Previous measurements of the triple-product asymmetries in \Bs decays from the \lhcb and \cdf
collaborations~\cite{LHCb-PAPER-2014-026,Aaltonen:2011rs} have shown no significant deviations from zero.

The \BsPP decay is a  $P\to VV$ decay, where $P$ denotes a pseudoscalar and $V$ a vector meson.
This gives rise to longitudinal and transverse polarisation of the final states with respect to their direction of flight in the \Bs reference frame, the fractions of which are denoted by $f_L$ and $f_T$, respectively.
In the heavy quark limit, $f_L$ is expected to be close to unity at tree level
due to the
%$V$--$A$ (vector-axial)
vector-axial structure of charged weak currents~\cite{Beneke:2006hg}.
This is found to be the case for tree-level \PB decays measured at the 
\bfactories~\cite{PhysRevLett.94.221804,PhysRevLett.98.051801,PhysRevD.78.092008, delAmoSanchez:2010mz,Abe:2004mq, Aubert:2006fs}.
However, the dynamics of penguin transitions are more complicated. 
Previously LHCb reported a value of  $f_L\equiv|A_0|^2 = 0.364 \pm 0.012$ in \BsPP decays~\cite{LHCb-PAPER-2014-026}. The measurement
is in agreement with predictions from QCD factorisation~\cite{Beneke:2006hg, PhysRevD.80.114026}.
%In the context
%of QCDf, $f_L\equiv|A_0|^2$ is predicted to be %$0.36^{+0.23}_{-0.18}$ for the \BsPP decay~\cite{PhysRevD.80.114026}. 
The observed value of $f_L$ is significantly larger than that seen in
the $\Bs \rightarrow \Kstarz \Kstarzb$ decay~\cite{LHCb-PAPER-2014-068, LHCb-PAPER-2017-048}.

In addition to the study of the \BsPP decay, a search for the as yet unobserved decay $\Bz\to\phi\phi$ is made. In the SM this is an OZI suppressed decay~\cite{Okubo:1963fa, *Iizuka:1966fk},  with an expected branching
fraction in the range $(0.1 - 3.0) \times 10^{-8}$~\cite{Lu:2005be, BarShalom:2002sv, Beneke:2006hg, Bartsch:2008ps}.  
However, the branching fraction can be enhanced, up to the $10^{-7}$ level, in extensions to the SM such as
supersymmetry with R-parity violation~\cite{BarShalom:2002sv}. 
The most recent experimental limit was determined to be $2.8\times10^{-8}$ at 90\,\% confidence level~\cite{LHCb-PAPER-2015-028}.

Measurements presented in this paper are based on $pp$ collision data
corresponding to an integrated luminosity of  $5.0\invfb$, 
collected with the \lhcb experiment at centre-of-mass energies $\sqrt{s}=7\tev$ in 2011, 8\tev in 2012,
and 13\tev from 2015 to 2016. This paper reports a time-dependent analysis of \BsPP decays, where the $\phi$ meson is reconstructed in the $\Kp\Km$ final state, that measures the \CP-violating phase, $\phi_s^{s\bar{s}s}$, and the parameter $|\lambda|$, that is related to the direct \CP violation. Results on helicity-dependent weak phases are also presented,
along with helicity amplitudes describing the $P\to VV$ transition and strong phases of the amplitudes. In addition, triple-product asymmetries for this decay are presented. The analysis also includes a search for the decay $B^0\rightarrow \phi\phi$.
Results presented here supersede the previous measurements
based on data collected in 2011 and 2012~\cite{LHCb-PAPER-2014-026}.

\section{Detector description}
\label{sec:Detector}

The \lhcb detector~\cite{LHCb-DP-2012-002,LHCb-DP-2014-002} is a single-arm forward
spectrometer covering the \mbox{pseudorapidity} range $2<\eta <5$,
designed for the study of particles containing \bquark or \cquark
quarks. The detector includes a high-precision tracking system
consisting of a silicon-strip vertex detector surrounding the $pp$
interaction region~\cite{LHCb-DP-2014-001}, a large-area silicon-strip detector located
upstream of a dipole magnet with a bending power of about
$4{\mathrm{\,Tm}}$, and three stations of silicon-strip detectors and straw
drift tubes~\cite{LHCb-DP-2017-001} placed downstream of the magnet.
The tracking system provides a measurement of the momentum, \ptot, of charged particles with
a relative uncertainty that varies from 0.5\% at low momentum to 1.0\% at 200\gevc.
The minimum distance of a track to a primary vertex (PV), the impact parameter (IP), 
is measured with a resolution of $(15+29/\pt)\mum$,
where \pt is the component of the momentum transverse to the beam, in\,\gevc.
Different types of charged hadrons are distinguished using information
from two ring-imaging Cherenkov detectors~\cite{LHCb-DP-2012-003}. 
Photons, electrons and hadrons are identified by a calorimeter system consisting of
scintillating-pad and preshower detectors, an electromagnetic
% calorimeter
and a hadronic calorimeter. Muons are identified by a
system composed of alternating layers of iron and multiwire
proportional chambers~\cite{LHCb-DP-2012-002}.

The online event selection is performed by a trigger, 
which consists of a hardware stage, based on information from the calorimeter and muon
systems, followed by a software stage, which applies a full event
reconstruction.
At the hardware trigger stage, events are required to contain a muon with high \pt or a hadron,
photon or electron with high transverse energy in the calorimeters.
In the software trigger, \BsPP candidates are selected 
either by identifying events containing a pair of oppositely charged kaons 
with an invariant mass within 30\mevcc of the known $\phi$ meson mass, $m_{\phi} = 1019.5\mevcc$~\cite{PDG2018}, or by using a topological \bquark-hadron trigger.
This topological trigger requires a three-track
secondary vertex with a large sum of the \pt of
the charged particles and significant displacement from the PV. 
At least one charged particle should have $\pt >
1.7\gevc$ and \chisqip with respect to any
primary vertex greater than 16, where \chisqip is defined as the
difference in \chisq of a given PV fitted with and
without the considered track.
A multivariate algorithm~\cite{BBDT} is used for
the identification of secondary vertices consistent with the decay
of a \bquark hadron.

Simulation samples are used to optimise the signal candidate selection, to derive the angular acceptance and the correction to the decay-time acceptance. 
In the simulation, $pp$ collisions are generated using
\pythia~\cite{Sjostrand:2006za,*Sjostrand:2007gs} with a specific \lhcb
configuration~\cite{LHCb-PROC-2011-006}.  Decays of hadronic particles
are described by \evtgen~\cite{Lange:2001uf}, in which final-state
radiation is generated using \photos~\cite{Golonka:2005pn}. The
interaction of the generated particles with the detector and its
response are implemented using the \geant
toolkit~\cite{Allison:2006ve, *Agostinelli:2002hh}, as described in
Ref.~\cite{LHCb-PROC-2011-006}.

\section{Selection and mass model}
\label{sec:selection}

For decay-time-dependent measurements and the $T$-odd asymmetries presented in this paper, the previously analysed data collected in 2011 and 2012~\cite{LHCb-PAPER-2014-026}
is supplemented with the additional data taken in 2015 and 2016, to which the selection
described below is applied.
For the case of the $\Bd\to\phi\phi$ search, a wider invariant-mass window
is required, along with more stringent background rejection requirements.

% Stripping decscription
Events passing the trigger are required to
satisfy loose criteria on the fit quality of the four-kaon vertex, the \chisqip of each
track, the transverse momentum of each particle, and the product of the transverse 
momenta of the two $\phi$ candidates. In addition, the reconstructed mass of 
the $\phi$ candidates is required to be within 25\mevcc of the known $\phi$ mass~\cite{PDG2018}.

% MLP
In order to separate further the \BsPP signal candidates from the background, a multilayer perceptron (MLP)~\cite{hastie01statisticallearning}
is used. To train the MLP, simulated
\BsPP candidates satisfying the same requirements as the data candidates are used as a proxy for signal,
whereas the four-kaon invariant-mass sidebands from data are used as a proxy for background. 
%A signal mass region is defined to be within 120\mevcc of the known \Bs mass, $m_{\Bs}= 5366.89\mevcc$~\cite{PDG2018}. 
The invariant-mass
sidebands are defined to be inside the region \mbox{$120<|m_{\Kp\Km\Kp\Km}-m_{\Bs}|<180\mevcc$}, where $m_{\Kp\Km\Kp\Km}$ is the four-kaon invariant mass.
Separate MLP classifiers are trained for each data taking period.
The variables used in the MLP comprise the minimum and the maximum
\pt and $\eta$ of the kaon and $\phi$ candidates, the \pt and $\eta$ of the \Bs candidate, 
the quality of the four-kaon vertex fit, and
the cosine of the angle between the momentum of the \Bs and the direction of flight
from the PV to the \Bs decay vertex, where the PV is chosen as that with the smallest
impact parameter $\chi^2$ with respect to the \Bs candidate.
For measurements of \CP violation, the requirement on each MLP is chosen to maximise $N_{\rm S}/\sqrt{N_{\rm S}+N_{\rm B}}$, where
${N_{\rm S}\,(N_{\rm B})}$ represents the expected signal and background yields in the signal region, defined as $m_{\Bs} \pm 3 \sigma$, where $m_{\Bs}$ is the known \Bs mass~\cite{PDG2018}. 
The signal yield is estimated using simulation,
whereas the number of background candidates is estimated from the data sidebands.
For the search of the $\Bd\rightarrow \phi\phi$ decay, the figure of merit is
chosen to maximise $\varepsilon/(a/2+\sqrt{N_{\rm B}})$~\cite{Punzi:2003bu}, where $a=3$ corresponds to the desired significance, and $\varepsilon$ is the signal efficiency,
determined from simulation.
%, and $3/2$ corresponds to a three standard deviation significance test.
This figure of merit does not depend on the unknown $\Bd\rightarrow \phi\phi$ decay rate.

The presence of peaking backgrounds is studied using simulation. The decay modes considered include $\Bd\to\phi\Kstarz$, 
$\Lb\to\phi \proton\Km$, $\Bd\to\phi\pip\pim$ and $\Bp\to\phi\Kp$, where the last decay mode could contribute if an extra kaon track is added. The $\Bd\to\phi\pip\pim$ and $\Bp\to\phi\Kp$ decays do not contribute significantly. 
%Only the last two, which are the results of a misidentification of a pion or proton as a kaon, respectively, are found to contribute significantly. 
The $\Bd\to\phi\Kstarz$ decay, resulting from a misidentification of a pion as a kaon, is vetoed by rejecting candidates which simultaneously 
have $\Kp\pim$($\Kp\Km\Kp\pim$) invariant masses within 50 (30)\mevcc
of the known \Kstarz (\Bd) masses. The $\Kp\pim$ and $\Kp\Km\Kp\pim$ invariant masses are computed by taking the kaon with the highest probability of being misidentified as a pion and assigning it the pion mass. 
These vetoes reduce the number of $\Bd\to\phi\Kstarz$ candidates to a negligible level.
Similarly, the number of $\Lb\to\phi \proton\Km$ decays, resulting from a misidentification of a proton as a kaon, is estimated from
data by assigning the proton mass to the final-state particle that has the largest probability to be a misidentified proton based on the particle-identification information.
This method yields $241\pm30$ $\Lb\to\phi \proton\Km$ decays in the total data set.

\begin{figure}[t]
\centering
\includegraphics[width=0.49\textwidth]{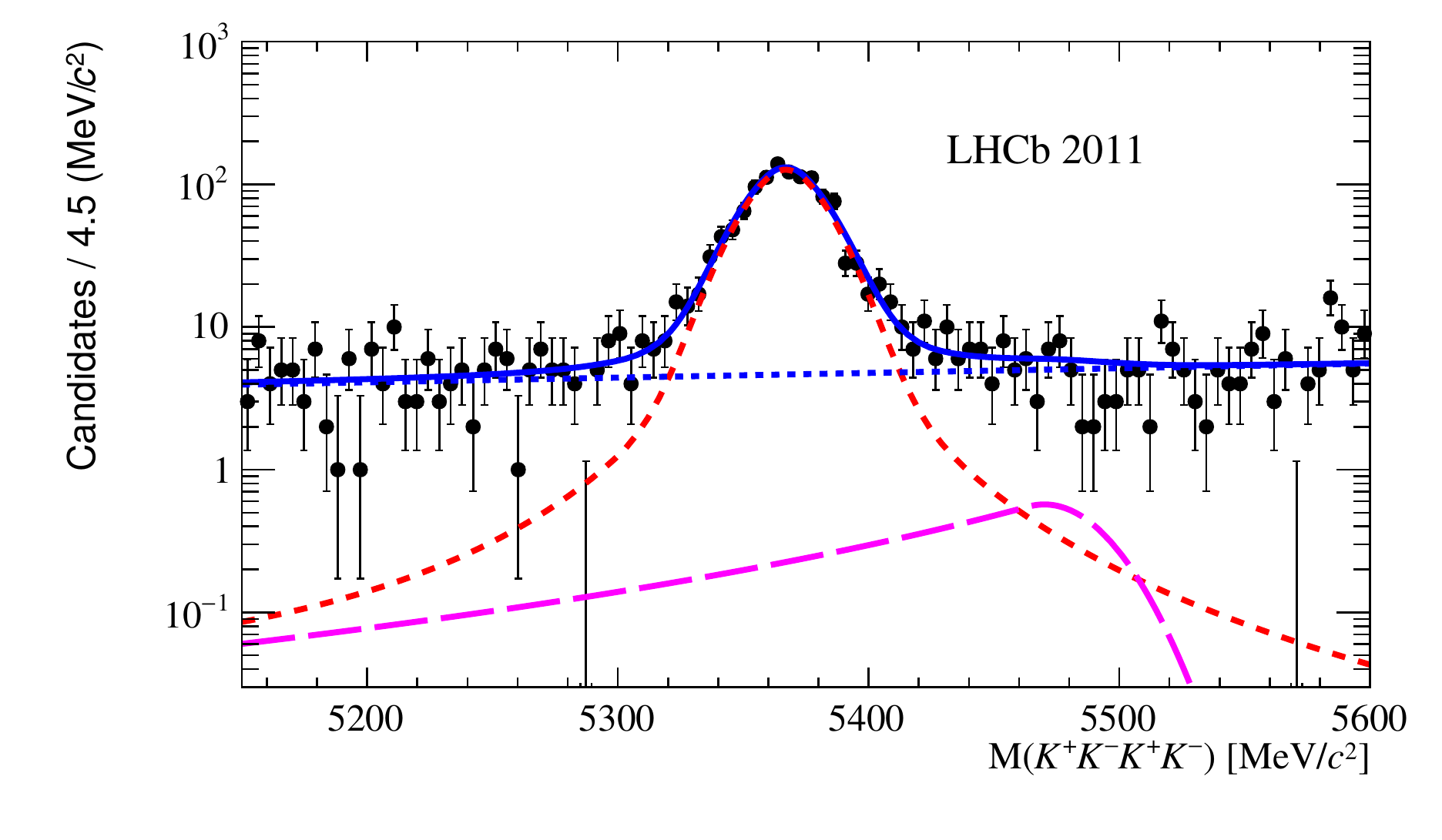}
\includegraphics[width=0.49\textwidth]{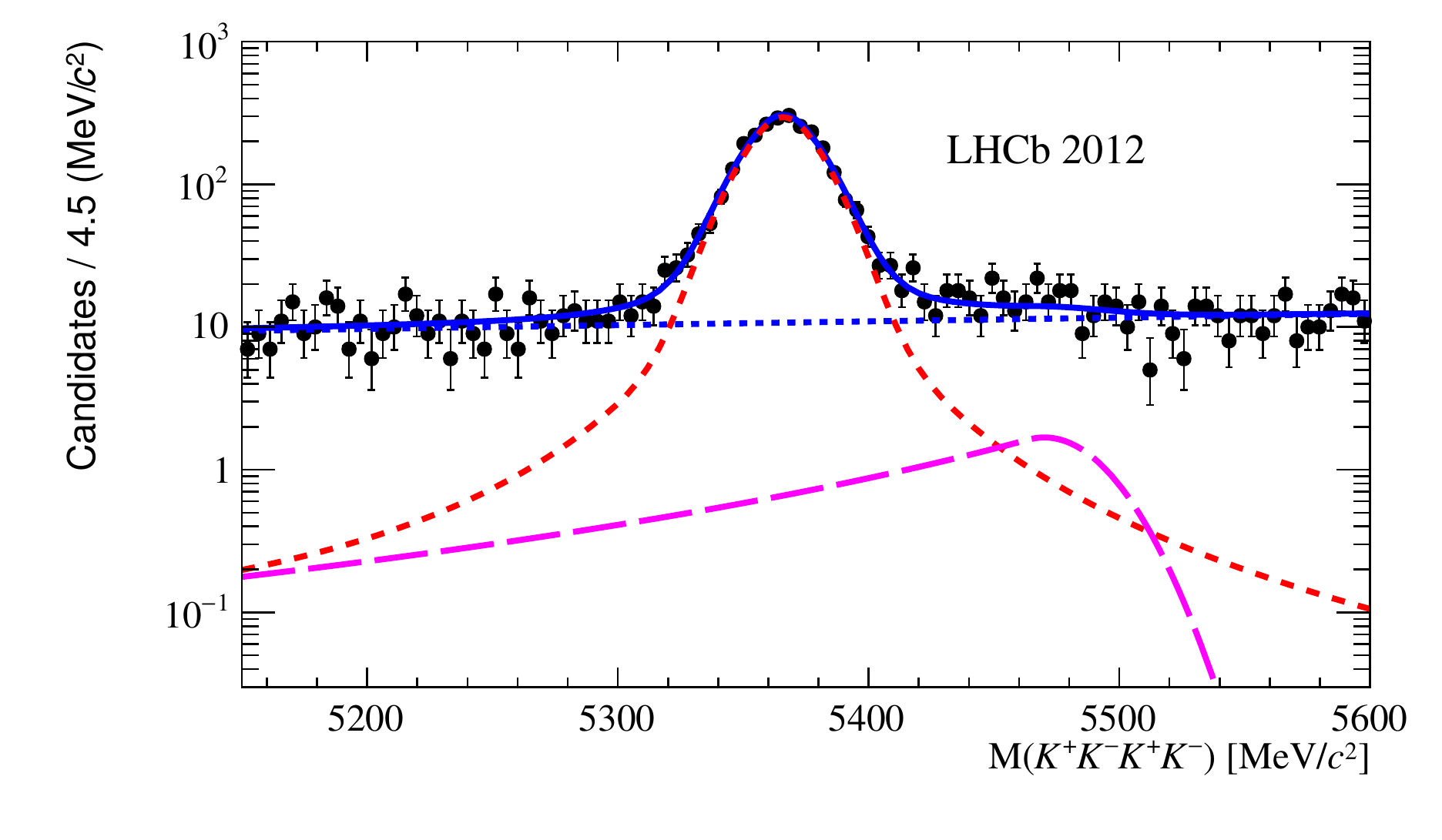} \\
\includegraphics[width=0.49\textwidth]{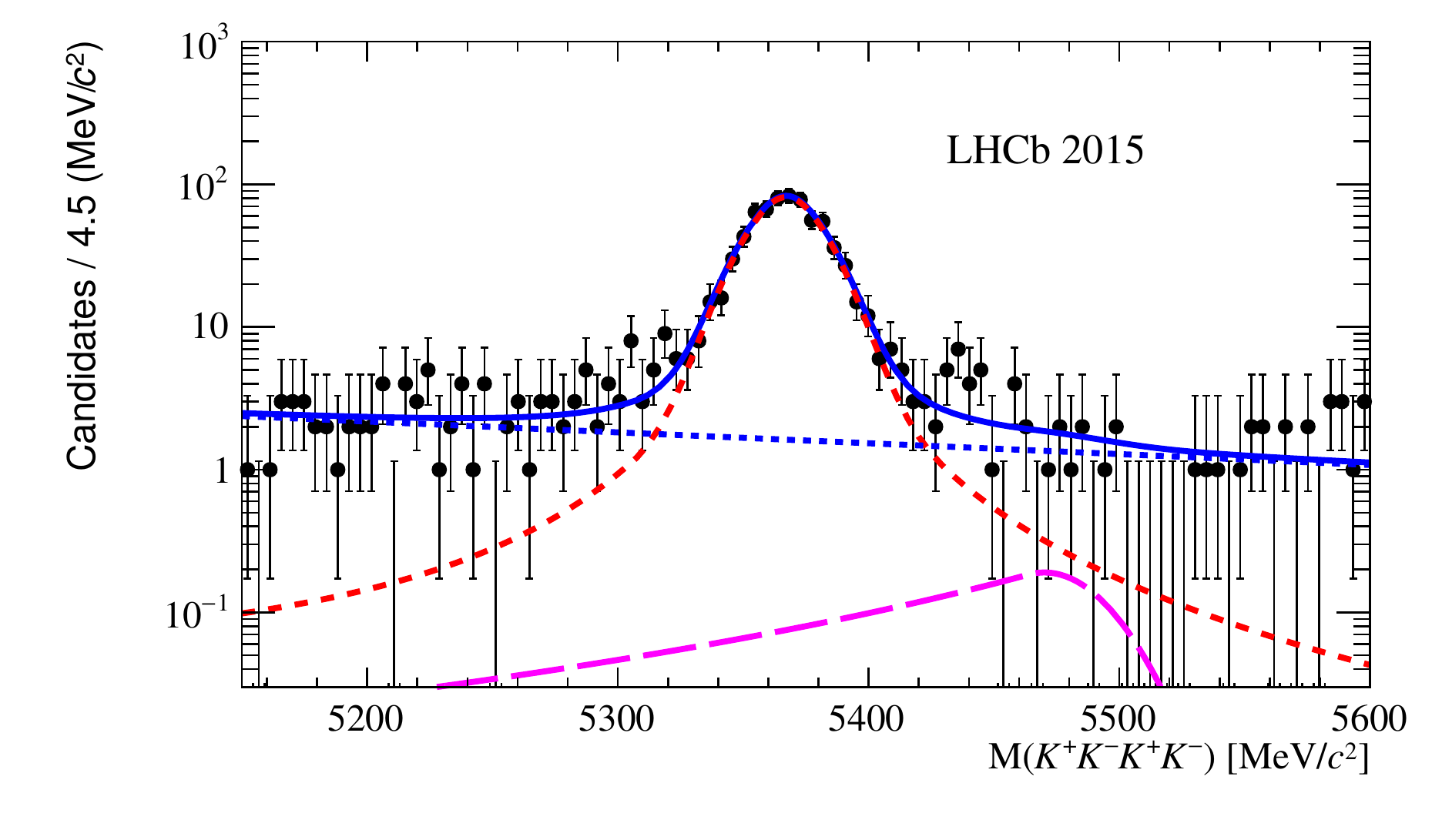}
\includegraphics[width=0.49\textwidth]{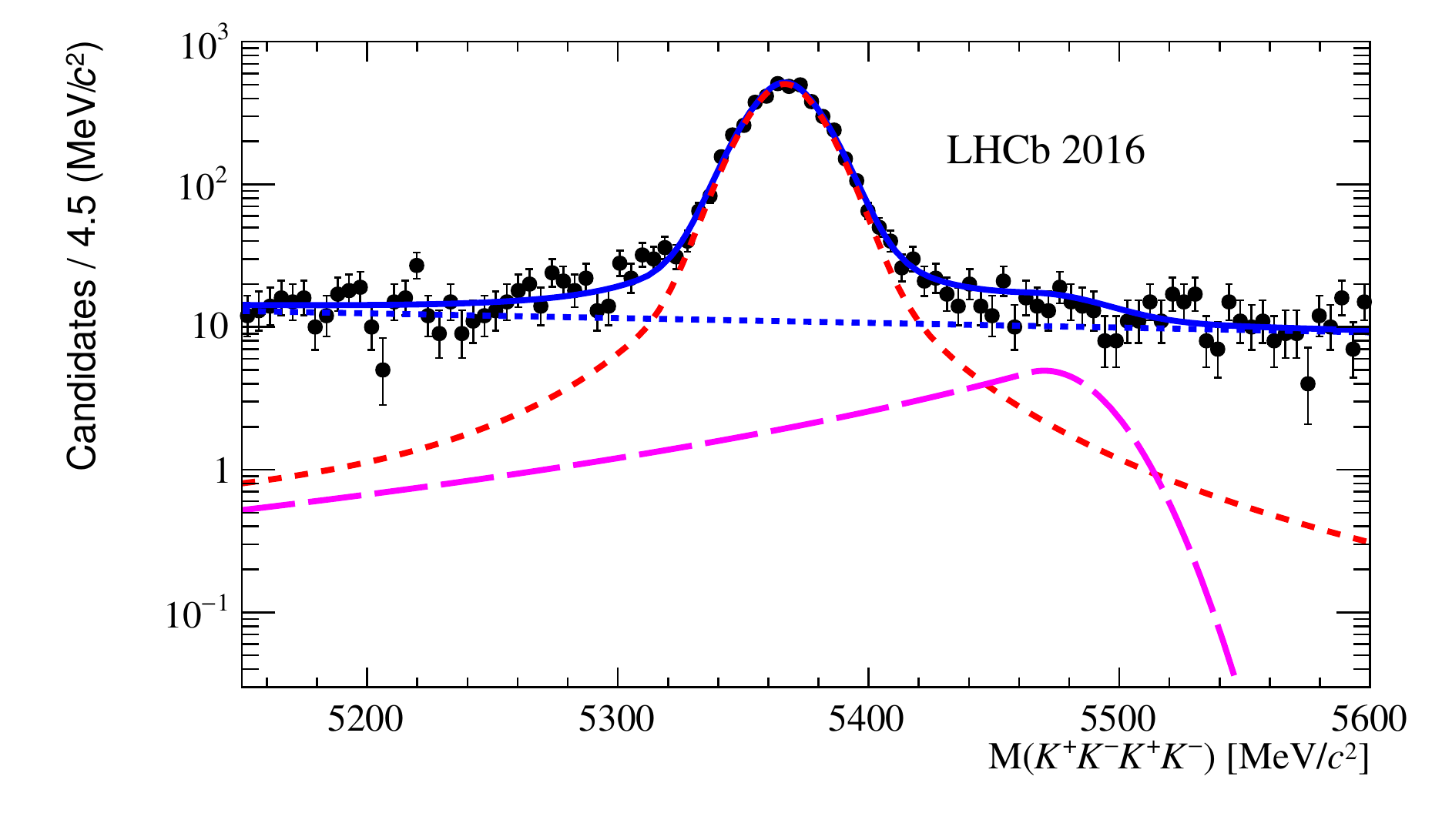}
\caption{A fit to the four-kaon mass for the (top left) 2011, (top right) 2012, (bottom left) 2015 and (bottom right) 2016 data sets,
which are represented by the black points. Also shown are the results of the total fit
(blue solid line), with the \BsPP (red dashed),
the $\Lb\to\phi \proton\Km$ (magenta long dashed), and the combinatorial (blue short dashed) fit components.
%The normalised residuals (pulls) with respect to total fits are given in the upper panels.
  }
\label{fig:massPlots}
\end{figure}

In order to determine the \BsPP yield in the final data sample, 
the four-kaon invariant-mass distributions are fitted with the sum of the following components:
a \BsPP signal model, which comprises the sum of a Crystal Ball~\cite{Skwarnicki:1986xj} and a Student's t-function; the peaking background contribution modelled by a Crystal Ball function, with the shape
parameters fixed to the values obtained from a fit to simulated events, and the combinatorial background component, described using
an exponential function. The yield of the $\Lb\to\phi\proton\Km$ peaking background contribution is fixed to the number previously stated.
Once the MLP requirements are imposed, an unbinned extended maximum-likelihood 
fit to the four-kaon invariant mass gives a total yield 
of $8843\pm102$ \BsPP decays and $2813\pm67$ combinatorial background candidates in the total data set.
The fits to the four-kaon invariant-mass distributions, after the selection optimised for the \CP-violation measurement, separately for each data taking year, are shown in Fig.~\ref{fig:massPlots}.

\section{Formalism}
\label{sec:phenom}

\begin{figure}[t]
\setlength{\unitlength}{1mm}
  \centering
  \begin{picture}(140,60) %90
    \put(0,-1){
      \includegraphics*[width=140mm,%height=100mm,%
      ]{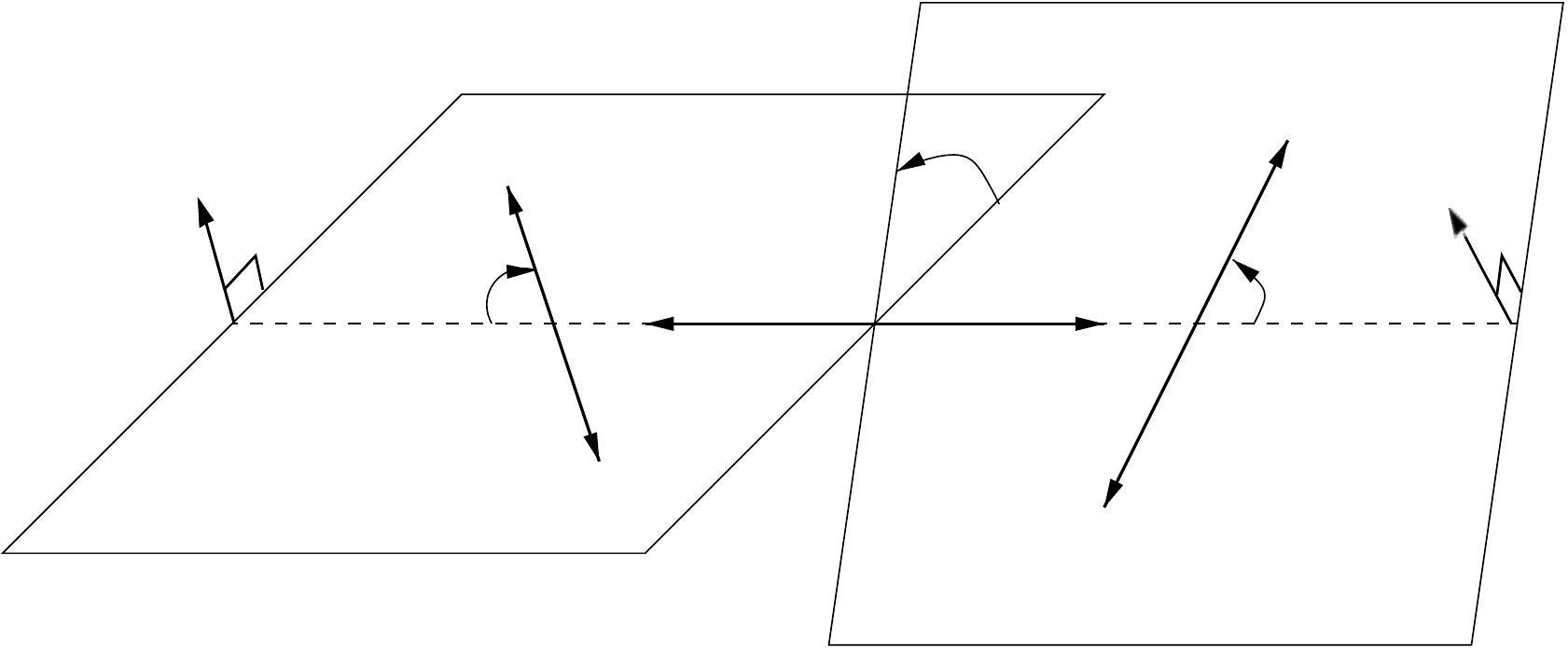}
    }
      \put(15,42){$\hat{n}_{V_1}$}
      \put(128,41){$\hat{n}_{V_2}$}
      \put(80,22){$B^0_{s}$}
      \put(83,37){$\Phi$}
      \put(117,30){$\theta_{2}$}
      \put(40,30){$\theta_{1}$}
      \put(36,16){$K^-$}
      \put(48,40){$K^+$}
      \put(105,15){$K^-$}
      \put(117,42){$K^+$}
      \put(60,30){$\phi_1$}
      \put(95,30){$\phi_2$}   
  \end{picture}
\caption{\small Decay angles for the $\Bs \rightarrow \phi \phi$ decay, where  
$\theta_{1,2}$ is the angle between the $K^+$ momentum in the $\phi_{1,2}$ meson rest frame and
the $\phi_{1,2}$ momentum in the \Bs rest frame, $\Phi$ is the angle between the two $\phi$ meson
decay planes and $\hat{n}_{V_{1,2}}$ is the unit vector normal to the decay plane of the $\phi_{1,2}$ meson.}
\label{fig:angles}
\end{figure}
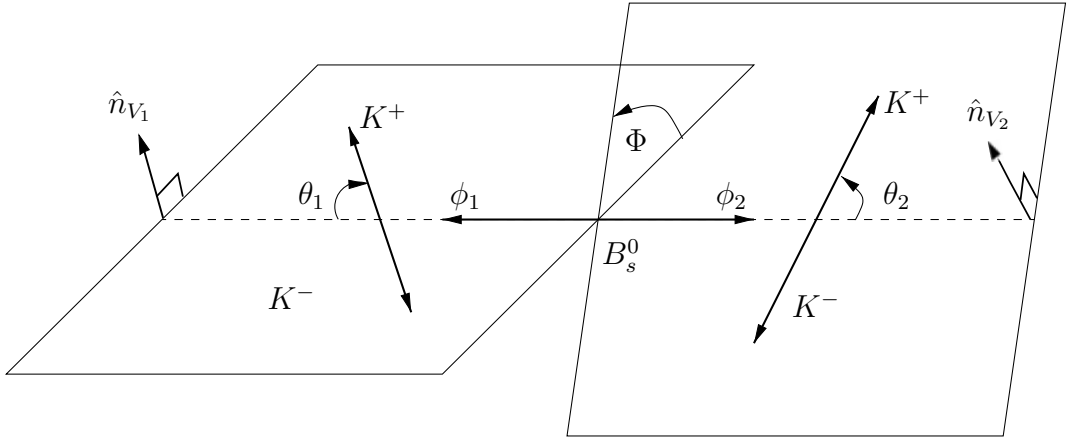

The final state of the \BsPP decay comprises a mixture of \CP eigenstates,
which are disentangled by means of an angular analysis
in the helicity basis. In this basis, the decay is described by three angles, $\theta_1$, $\theta_2$ and $\phi$, defined in Fig.~\ref{fig:angles}.

\subsection{Decay-time-dependent model}
\label{sec:modelTD}

As discussed in Sec.~\ref{sec:Introduction},
the \BsPP decay is a  $P\to VV$ decay.
However, due to the proximity of the $\phi$ 
resonance to the scalar $f_0(980)$ resonance, 
there are irreducible contributions to the four-kaon mass spectrum from $P\to V\kern-0.1em S$ ($S$-wave) and $P\to S\kern-0.1em S$ (double $S$-wave) processes, where $S$ denotes 
a scalar meson, or a nonresonant pair of kaons.
Thus, the total amplitude is a coherent sum of $P$-, $S$-, and double $S$-wave processes, and is
%The $S$-wave, referring to the $P\to VS$ and $P\to SS$ processes, is 
modelled
by making use of the different dependence on the helicity angles associated with these terms,
where the helicity angles are defined in Fig.~\ref{fig:angles}.
A randomised choice is made for which $\phi$ meson is used to determine $\theta_1$ and which is
used to determine $\theta_2$.
The total amplitude ($\mathcal{A}$) containing the $P$-, $S$-, and double $S$-wave components as
a function of time, $t$, can be written as~\cite{Bhattacharya:2013sga}
\begin{align}
\mathcal{A}(t,\theta_1,\theta_2,\Phi) &= A_0(t)\cos\theta_1\cos\theta_2 
+\frac{A_\parallel(t)}{\sqrt{2}}\sin\theta_1\sin\theta_2\cos\Phi \nonumber \\
&+i\frac{A_\perp(t)}{\sqrt{2}}\sin\theta_1\sin\theta_2\sin\Phi
+\frac{A_S(t)}{\sqrt{3}}(\cos\theta_1 + \cos\theta_2)
+\frac{A_{SS}(t)}{3},
\end{align}
where $A_0$, $A_\parallel$, and $A_\perp$ are the \CP-even longitudinal, \CP-even parallel, and \CP-odd perpendicular
polarisations of the \BsPP decay. The $P\to V\kern-0.1em S$ and $P\to S\kern-0.1em S$ processes are described by the $A_S$ and $A_{SS}$ amplitudes, respectively, where $P\to V\kern-0.1em S$ is \CP-odd and $P\to S\kern-0.1em S$ is \CP-even.
The resulting differential decay rate is proportional to the square of the total amplitude and consists of 15 terms~\cite{Bhattacharya:2013sga}
\begin{align}
\frac{\deriv \Gamma}{\deriv t\,\deriv \cos\theta_1\,\deriv \cos\theta_2 \,\deriv\Phi} \propto |\mathcal{A}(t,\theta_1,\theta_2,\Phi)|^2
= \frac{1}{4}\sum_{i=1}^{15} K_i (t) f_i (\theta_1,\theta_2,\Phi),
\label{eq:pdf}
\end{align}
where the $f_i$ terms are functions of the angular variables and the time-dependence is contained in

\begin{align}
K_i(t)=N_ie^{-\Gs t} \left[ 
a_i \cosh\left(\frac{1}{2}\DGs t\right) + b_i \sinh\left(\frac{1}{2}\DGs t\right) + c_i \cos(\dms t) + d_i\sin(\dms t)\right]\kern-0.25em .
\label{eq:pdfK}
\end{align}
The coefficients $N_i, a_i, b_i, c_i$ and $d_i$, which are functions of the \CP observables, are defined in Appendix~\ref{app:terms}.
$\DGs \equiv \Gamma_{\rm L} - \Gamma_{\rm H}$ is the decay-width difference between the
light and heavy \Bs mass eigenstates,
 $\Gs \equiv (\Gamma_{\rm L }+ \Gamma_{\rm H})/2$ is the average decay width, 
  and $\dms$ is the \Bs-\Bsb oscillation frequency. 
The differential decay rate for a \Bsb meson produced at $t=0$ is obtained
by changing the sign of the $c_i$ and $d_i$ coefficients.
The amplitudes of helicity state $k$ are expressed as
\begin{equation}
A_k(t) = |A_k|e^{i\delta_k}\left( g_+(t) + \eta_k |\lambda_k| e^{-i\phi_{s,k}} g_-(t) \right),
\end{equation}
where $g_+(t)$ and $g_-(t)$ describe the time evolution of \Bs and \Bsb mesons, respectively. 
\CP violation is parameterised through
\begin{equation}
\frac{q}{p}\frac{\bar{A}_k}{A_k} = \eta_k |\lambda_k|e^{-i\phi_{s,k}}.
\label{eq:cpvpol}
\end{equation}
where, $q$ and $p$ relate the light and heavy mass eigenstates to the
flavour eigenstates and $\eta_k$ is the \CP eigenvalue of the polarisation being considered. 
Defining the amplitude in this way leads to the forms of $N_i, a_i, b_i, c_i$ and $d_i$, listed in Table~\ref{tab:terms} (Appendix~\ref{app:terms}).
The \CP-violating asymmetry in \Bs mixing, which can be characterised by the semileptonic asymmetry, $a_{\mathrm{sl}}^s$ is small~\cite{LHCB-PAPER-2016-013}. Thus, to good approximation $|q/p|=1$, and $|\lambda_k|$ quantifies the level of \CP violation in the decay.  
%Under the assumption that $|q/p|=1$, $|\lambda_k|$ measures \CP violation in decay.
Two different fit configurations are performed, one in which the 
\CP-violation parameters are assumed to be helicity independent and the other in
which \CP-violation parameters are allowed to differ as a function of helicity.
The helicity independent fit assumes one \CP-violating phase, \phisPP, which takes the place of all $\phi_{s,k}$ contained in the coefficients of Appendix~\ref{app:terms}, and likewise one parameter that describes direct \CP violation, $|\lambda|$, which takes the place of all $\lambda_k$ coefficients.
Due to the small sample size, the number of degrees of freedom is reduced for the
case of the helicity-dependent \CP-violation fit. This involves assuming \CP conservation
for the case of the direct \CP-violation parameters, $\lambda = 1$, and also for the phase of the
longitudinal polarisation, $\phi_{s,0}^{s\bar{s}s} = 0$. The longitudinal polarisation has been theoretically calculated as close to zero
in the \BsPP decay~\cite{Bartsch:2008ps}.

The \phisPP and $|\lambda|$ parameters are measured with respect to contributions with the same flavour content as the $\phi$ meson, \ie~\ssbar. Regarding the $S$-wave and double $S$-wave terms, the impact of the non-\ssbar component
of the $\phi$ wavefunction is negligible in this analysis.
%The association of \phisPP and $|\lambda|$ with $S$-wave and double $S$-wave terms implies that these consist solely of contributions
%with the same flavour content as the $\phi$ meson, \ie ~ \ssbar. The impact of the non-\ssbar component
%of the $\phi$ wavefunction is negligible in this analysis.

%In Table~\ref{tab:terms},
%$\delta_S$ and $\delta_{SS}$ are the strong phases of the $P\to V\kern-0.1em S$ and $P\to S\kern-0.1em S$ processes, respectively.
%The $P$-wave strong phases are defined to be $\delta_1\equiv\delta_\perp-\delta_\parallel$ and $\delta_2\equiv\delta_\perp-\delta_0$,
%with the notation $\delta_{2,1}\equiv\delta_2-\delta_1=\dpa-\dz$.
%
%
%

\subsection{Triple-product asymmetries}
\label{sec:modelTP}

Scalar triple products of three-momentum or spin vectors are odd under
time reversal, ~\T. Nonzero asymmetries for these observables can either be due to a CP-violating phase or from CP-conserving strong final-state interactions.
%a \CP-conserving phase and final-state interactions. 
Four-body final states give rise to three independent momentum vectors in the rest frame of the decaying \Bs
meson. For a detailed review of the phenomenology the reader is referred to Ref.~\cite{gronau}. 

%The two independent terms in the decay-time-dependent decay rate that contribute to a $T$-odd asymmetry
%are the $K_4(t)$ and $K_6(t)$ terms, defined in Eq.~\ref{eq:pdfK}. The triple products that allow access to these terms are
Two triple products can be defined:
\begin{eqnarray}
\sin \Phi = (\hat{n}_{V_1} \times \hat{n}_{V_2}) \cdot \hat{p}_{V_1}, \\
\sin 2\Phi = 2(\hat{n}_{V_1} \cdot \hat{n}_{V_2})(\hat{n}_{V_1} \times \hat{n}_{V_2}) \cdot \hat{p}_{V_1},
\end{eqnarray}
where $\hat{n}_{V_i}$ ($i = 1,2$) is a unit vector perpendicular to the vector meson ($V_i$) decay 
plane and $\hat{p}_{V_1}$ is a unit vector in the direction of $V_1$ in the \Bs rest frame, defined in
Fig.~\ref{fig:angles}.
This then provides a method of probing \CP violation without the need to measure the decay time 
or the initial flavour of the \Bs meson.
It should be noted, that while the observation of nonzero triple-product asymmetries implies \CP violation or final-state interactions
(in the case of \Bs meson decays), measurements of triple-product asymmetries consistent with zero
do not rule out the presence of \CP-violating effects, as the size of the asymmetry also depends on the differences between the strong phases~\cite{gronau}.

In the $\BsPP$ decay, two triple products are defined as $U \equiv \sin\Phi\cos\Phi$
and $V \equiv \sin(\pm \Phi)$ where the positive sign is taken 
if $\cos \theta_1 \cos \theta_2 \geq 0$ and the negative sign
otherwise~\cite{gronau}. The \T-odd asymmetry corresponding to the $U$ observable, $A_U$, is defined as the normalised difference between the number of decays
with positive and negative values of $\sin\Phi\cos\Phi$,
\begin{align}
%A_U \equiv \frac{\Gamma(\sin 2 \Phi > 0) - \Gamma(\sin 2 \Phi < 0)}{\Gamma(\sin 2 \Phi > 0) + \Gamma(\sin 2 \Phi < 0)} \propto \int^\infty_0 \Im (A_\perp(t) A_\parallel^*(t) + \bar{A}_\perp(t) \bar{A}_\parallel^*(t)) \deriv t\, .
A_U \equiv \frac{\Gamma(U > 0) - \Gamma(U < 0)}{\Gamma(U > 0) + \Gamma(U < 0)} \propto \int^\infty_0 \Im \left( A_\perp(t) A_\parallel^*(t) + \bar{A}_\perp(t) \bar{A}_\parallel^*(t) \right) \deriv t.
\end{align}
Similarly, $A_V$ is defined as
\begin{eqnarray}
%A_V \equiv \frac{\Gamma({\rm sign}(\cos\theta_1\cos\theta_2)\sin \Phi > 0) - \Gamma({\rm sign}(\cos\theta_1\cos\theta_2)\sin \Phi < 0)}{\Gamma({\rm sign}(\cos\theta_1\cos\theta_2)\sin \Phi > 0) + \Gamma({\rm sign}(\cos\theta_1\cos\theta_2)\sin \Phi < 0)} \nonumber \\ \propto \int^\infty_0 {\Im} (A_\perp(t) A_0^*(t) + \bar{A}_\perp(t) \bar{A}_0^*(t)) \deriv t\, .
A_V \equiv \frac{\Gamma(V > 0) - \Gamma(V < 0)}{\Gamma(V > 0) + \Gamma(V < 0)} \propto \int^\infty_0 {\Im} \left( A_\perp(t) A_0^*(t) + \bar{A}_\perp(t) \bar{A}_0^*(t)\right) \deriv t.
\end{eqnarray}
Here, $A_\perp$, $A_\parallel$ and $A_0$ correspond to the three transversity amplitudes.
The determination of the triple-product asymmetries is then reduced to a
simple counting experiment.
Comparing these formulae with Eq.~\ref{eq:pdfK} and Appendix~\ref{app:terms} it can be seen that the
triple products are related to the $K_4(t)$ and $K_6(t)$ terms in the decay amplitude.
%that does not require tagging or a time
%dependent analysis.
%

\section{Decay-time resolution}
\label{sec:DTR}

The sensitivity to \phisPP is affected by the accuracy of the measured decay time.
In order to resolve the fast \Bs-\Bsb oscillations, it is necessary to have a decay-time
resolution that is much smaller than the oscillation period.
To account for the resolution of the measured decay-time distribution, all decay-time-dependent terms
are convolved with a Gaussian function, with width $\sigma^t_i$ that
is estimated for each candidate, $i$, based upon the uncertainty obtained from the vertex and kinematic fit~\cite{Hulsbergen:2005pu}.

In order to apply a candidate-dependent resolution model during fitting, the estimated per-event decay time uncertainty is needed. This is calibrated using the fact the decay time resolution for the \BsPP mode is dominated by the secondary vertex resolution. A sample of good-quality tracks, which originate from the primary interaction vertex is selected. Due to the small opening angle of the kaons in the decay of a $\phi$ meson, it is sufficient to use a single prompt track and assign it the mass of a $\phi$ meson. When combining this with another pair of tracks, the invariant mass of the three-body combination is required to be within 250 \mevcc of the known \Bs mass.
That the decay-time resolution of the signal \Bs decays can be described well using three tracks has been validated using simulation. 

A linear function is then fitted to the distribution of $\sigma^t_i$ versus $\sigma_{\rm true}^t$, 
with parameters $q_0$ and $q_1$. Here, $\sigma_{\rm true}^t$ denotes the
difference between reconstructed decay time and the exact decay time of simulated signal.
%truth-matched events.
The per-event decay-time uncertainty used in the decay-time-dependent fit
is then calculated as $\sigma^{\rm cal}_i = q_0 + q_1 \sigma^t_i$.
Gaussian constraints are used to account for the uncertainties on the calibration parameters
in the decay-time-dependent fit.
The effective single-Gaussian decay-time resolution is found to be
between 41 and 44\fs, depending on the data-taking year, in agreement with the expectation from the simulation.

\section{Acceptances}
\label{sec:acc}

The \BsPP differential decay rate depends on the decay time and three helicity angles as shown in Eq.~\ref{eq:pdf}.
Good understanding of the efficiencies in these variables is required.
The decay-time and angular acceptances are assumed to factorise. Control channels show this assumption has a negligible systematic uncertainty on the physics parameters.

\subsection{Angular acceptance}
\label{sec:AA}

\begin{figure}[t]
\centering
\includegraphics[width=0.45\textwidth]{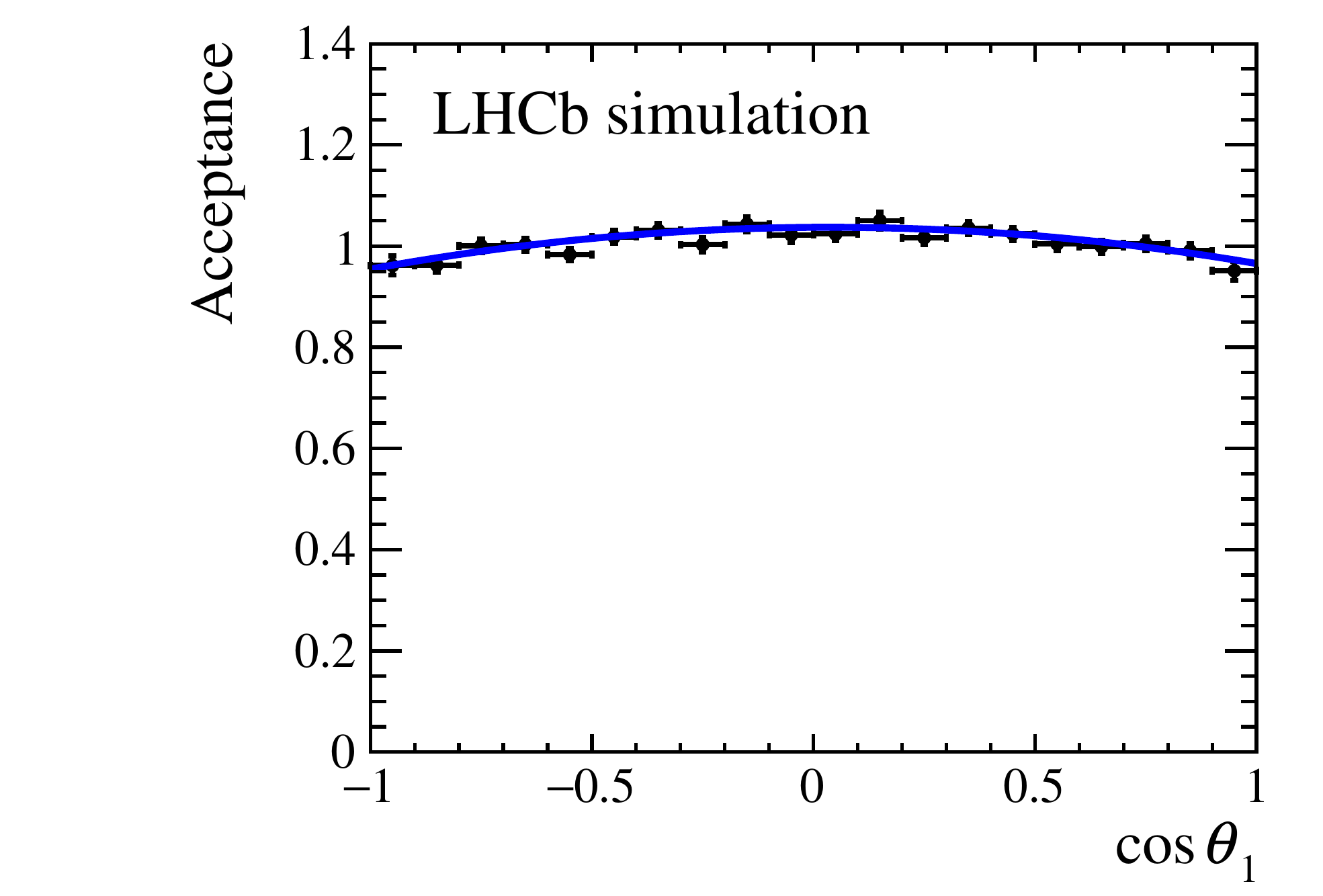}
\includegraphics[width=0.45\textwidth]{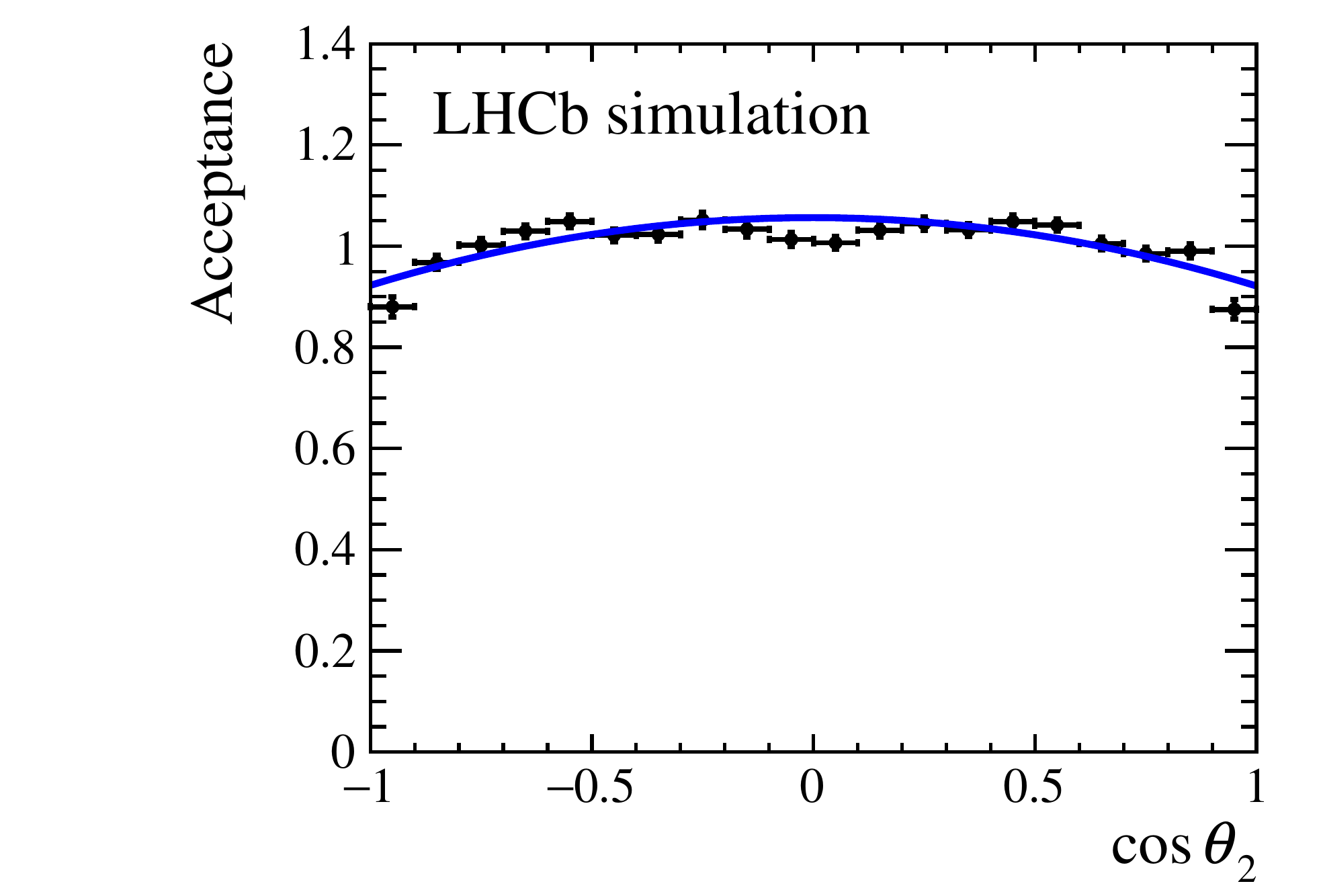}
\includegraphics[width=0.45\textwidth]{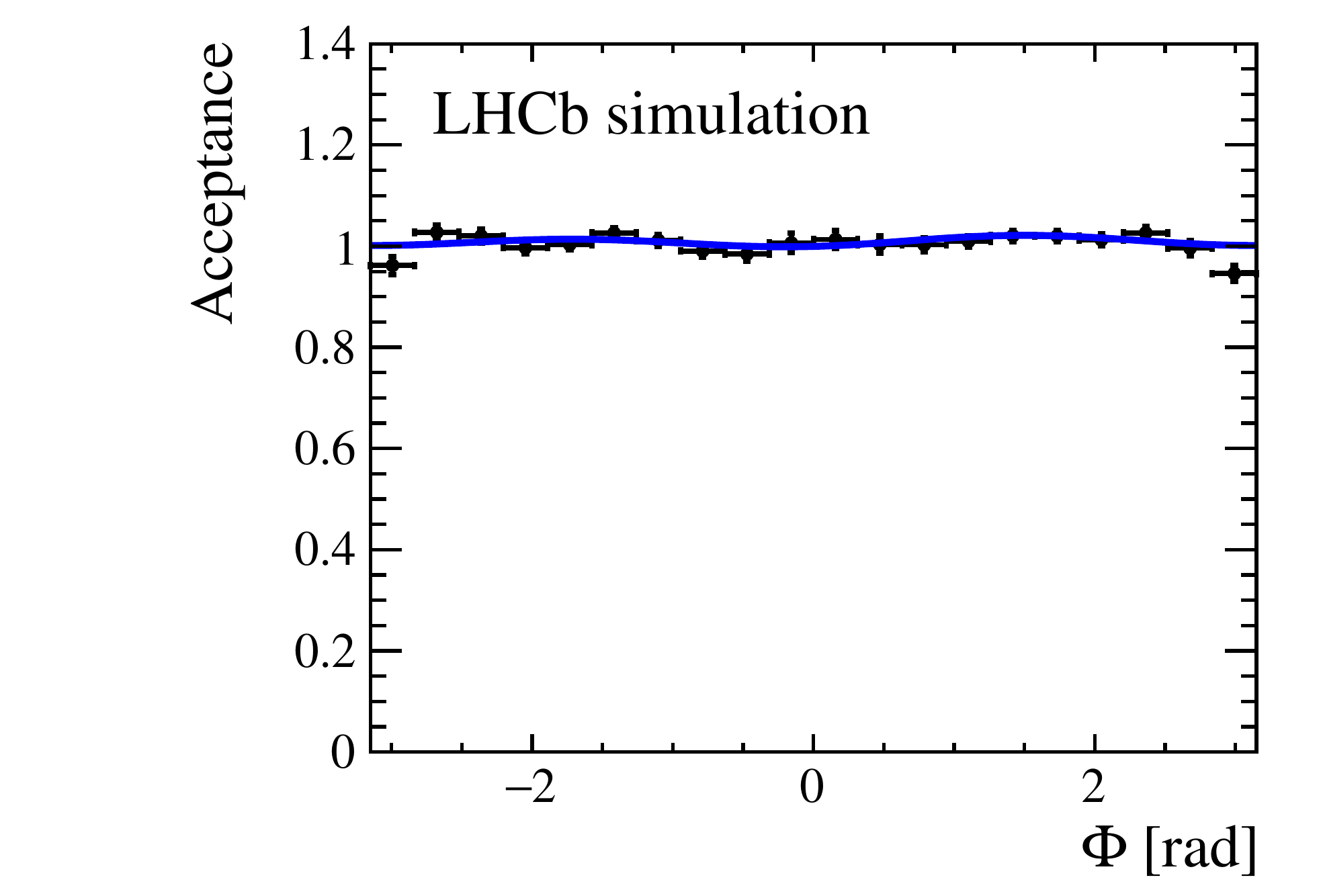}
\caption{
\small Angular acceptance normalised to the average obtained using simulated \BsPP decays (top-left) integrated over $\cos\theta_2$ and $\Phi$ as a function of $\cos\theta_1$, (top-right) integrated over
$\cos\theta_1$ and $\Phi$ as a function of $\cos\theta_2$, and (bottom) integrated over
$\cos\theta_1$ and $\cos\theta_2$ as a function of $\Phi$. Each figure includes the resulting fit curve.}
\label{fig:AngAcc}
\end{figure}

The geometry of the \lhcb detector and the momentum requirements imposed on the final-state particles
introduce distortions of the helicity angles, giving rise to acceptance effects. Simulated signal events, selected with the 
same criteria as those applied to data are used to determine these efficiency corrections.
The angular acceptances as a function of the three helicity angles are shown in Fig.~\ref{fig:AngAcc}.

The efficiency is parameterised in terms of the decay angles as

\begin{equation}
\label{eq:epilson}
\epsilon (\Omega) = \sum_{i,j,k} c_{ijk}P_i(\cos\theta_1)Y_{jk}(\cos\theta_2,\Phi ),
\end{equation}
where $\Omega$ depends on the decay angles, $\cos \theta_1$, $\cos \theta_2$ and $\phi$, the $c_{ijk}$ are coefficients, $P_i(\cos\theta_1)$
are Legendre polynomials, and $Y_{jk}(\cos\theta_2,\Phi )$ are spherical harmonics.
The procedure followed to calculate the coefficients 
is described in detail in Ref.~\cite{JVLthesis} and exploits the orthogonality of Legendre polynomials. The coefficients are given by
\begin{align}
c_{ijk}\equiv (j+1/2)\int d\Omega P_i(\cos\theta_1)Y_{jk}(\cos\theta_2,\Phi )\epsilon(\Omega).
\end{align}
This integral is calculated by means of a Monte Carlo technique, which reduces the integral
to a sum over the number of accepted simulated events ($N^{\rm obs}$)
\begin{align}
  c_{ijk}\propto (j+1/2) \frac{1}{N^{\rm obs}} \sum_{e=1}^{N^{\rm obs}} \frac{P_i(\cos\theta_{1,e})Y_{jk}(\cos\theta_{2,e},\Phi_e )}{P^{\rm gen}(\Omega_e|t_e)},
\end{align}
where $P^{\rm gen}$ is the probability density function (PDF) without acceptance
where the parameters are set to values used in the Monte Carlo generation.
In order to easily incorporate the angular acceptance, it is convenient to write angular functions of Eq.~\ref{eq:pdf} in the same basis as the efficiency parameterisation, \ie
\begin{align}
\label{eq:fk}
f_a(\cos\theta_1,\cos\theta_2,\Phi) = \sum_{ijkl}\kappa_{ijkl,a}P_{ij}(\cos\theta_1)Y_{kl}(\cos\theta_2,\Phi),
\end{align}
where $P_{ij}(\cos\theta_1)$ are the associated Legendre polynomials, $\kappa_{ijkl,a}$ are coefficients 
and $a$ numerates the 15 terms outlined earlier. 
The parameterisation for each angular
function is given in Table~\ref{tab:AngTermsPY}.

The normalisation of the angular component in the decay-time dependent fit occurs through the 15 integrals $\zeta_k= \int \epsilon(\Omega)f_k(\Omega)\mathrm{d}\Omega$, where $\epsilon(\Omega)$ is the efficiency as a function of the helicity angles as shown in Eq.~\ref{eq:epilson} and $f_k(\Omega)$ are the angular functions as defined in Eq.~\ref{eq:fk}. 

\begin{table}[bt]
\begin{center}
{
\caption{Angular coefficients, written in the same basis as the efficiency parameterisation.\label{tab:AngTermsPY}}
$
\scriptsize
%\[
\begin{array}{c|c|c}
i     & f_i\, \mathrm{in} P_{i\,j}Y_{k\,l}\,{\rm basis}               & f_i \\ \hline
1     & 8/9P_{00}Y_{00} + 16/9/\sqrt{5}P_{00}Y_{20} + 16/9P_{20}Y_{00} + 32/(9\sqrt{5})P_{20}Y_{20} & 4\cos^2\theta_1\cos^2\theta_2 \\\hline
2     & 8/9P_{00}Y_{00} - 8/9P_{00}Y_{20} - 8/9\sqrt{5}P_{20}Y_{00} + 8/(9\sqrt{5})P_{20}Y_{20} + (2/9)\sqrt{12/5}P_{22}Y_{22} & \sin^2\theta_1\sin^2\theta_2(1{+}\cos2\Phi) \\\hline
3     & 8/9P_{00}Y_{00} - 8/9P_{00}Y_{20} - 8/9\sqrt{5}P_{20}Y_{00} + 8/(9\sqrt{5})P_{20}Y_{20} - (2/9)\sqrt{12/5}P_{22}Y_{22} & \sin^2\theta_1\sin^2\theta_2(1{-}\cos2\Phi) \\ \hline
4     & -8/9\sqrt{3/5}P_{2,2}Y_{2,-2} & -2\sin^2\theta_1\sin^2\theta_2\sin 2\Phi \\\hline
5     &  8/9\sqrt{6/5}P_{2,1}Y_{2,1} & \sqrt{2}\sin2\theta_1\sin2\theta_2\cos\Phi \\\hline
6     & -8/9\sqrt{6/5}P_{2,1}Y_{2,1} & -\sqrt{2}\sin2\theta_1\sin2\theta_2\sin\Phi\\\hline
7     &  (8/9)P_{00}Y_{00}& \frac{4}{9} \\\hline
8     &  16/9P_{00}Y_{00} + 16/9/\sqrt{5}P_{00}Y_{20} + 16/9P_{20}Y{00} + 16\sqrt{3}/9P_{10}Y_{10} & \frac{4}{3}(\cos\theta_1+\cos\theta_2)^2 \\\hline
9     &  16\sqrt{3}/2P_{10}Y_{00} + 16/9P_{00}Y_{10} & \frac{8}{3\sqrt{3}}(\cos\theta_1+\cos\theta_2) \\\hline
10      & 16/(3\sqrt{3})P_{10}Y_{10} & \frac{8}{3}\cos\theta_1\cos\theta_2\\\hline
11      & (8/9)\sqrt{6}P_{11}Y_{11} & \frac{4\sqrt{2}}{3}\sin\theta_1\sin\theta_2\cos\Phi\\\hline
12      & (8/9)\sqrt{6}P_{11}Y_{1-1}& -\frac{4\sqrt{2}}{3}\sin\theta_1\sin\theta_2\sin\Phi\\\hline
13      & 16\sqrt{3}/9P_{10}Y_{00} + 16/9P_{00}Y_{10} + 32/9P_{20}Y{10} + 32/(9\sqrt{5})P_{20}Y_{20} & \begin{array}{c}\frac{8}{\sqrt{3}} \cos\theta_1\cos\theta_2\\\times (\cos\theta_1 + \cos\theta_2)\end{array}\\\hline
14      & (8/9)\sqrt{2/3}P_{21}Y_{11} + (24/9)\sqrt{2/15}P_{11}Y_{21} & \begin{array}{c}\frac{4\sqrt{2}}{\sqrt{3}} \sin\theta_1\sin\theta_2\\\times (\cos\theta_1 + \cos\theta_2)\cos\Phi\end{array}\\\hline
15      & - (8/9)\sqrt{2/3}P_{21}Y_{1-1} - (24/9)\sqrt{2/15}P_{11}Y_{2-1} & \begin{array}{c}-\frac{4\sqrt{2}}{3}\sin\theta_1\sin\theta_2\\\times (\cos\theta_1 + \cos\theta_2)\sin\Phi\end{array}\\%\hline
\end{array}
%\]
$
}
\end{center}
\end{table}

The angular acceptance is calculated correcting for the differences in kinematic
variables between data and simulation.
This includes differences in the MLP training variables that can affect acceptance 
corrections through correlations with the helicity angles.

The fit to determine the triple-product asymmetries assumes that the $U$ and $V$ observables
are symmetric in the acceptance corrections. Simulation is used to assign a 
systematic uncertainty related to this assumption.

\subsection{Decay-time acceptance}
\label{sec:DTA}

The impact-parameter requirements on the final-state particles efficiently suppress the background from
the numerous pions and kaons originating from the PV, but introduce a decay-time dependence in the
selection efficiency. 

The efficiency as a function of the decay time is taken from 
the \mbox{$\Bs\to\Dsm(\to \Kp\Km\pim) \pip$} decay, in the case of 
data taken between 2011 and 2012, and from the $\Bd\to\jpsi(\to\mup\mun)\Kstarz(\to\Kp\pim)$ 
decay in the case of data taken between 2015 and 2016.
The reason for the change in control channel is related to changes to the software-trigger selection between the two data-taking periods.
The decay-time acceptances of the control modes are weighted by a multivariate algorithm based on simulated kinematic and topological information, in order to match more closely those of the signal \BsPP decay.

\begin{figure}[t]
\centering
\includegraphics[width=0.45\textwidth]{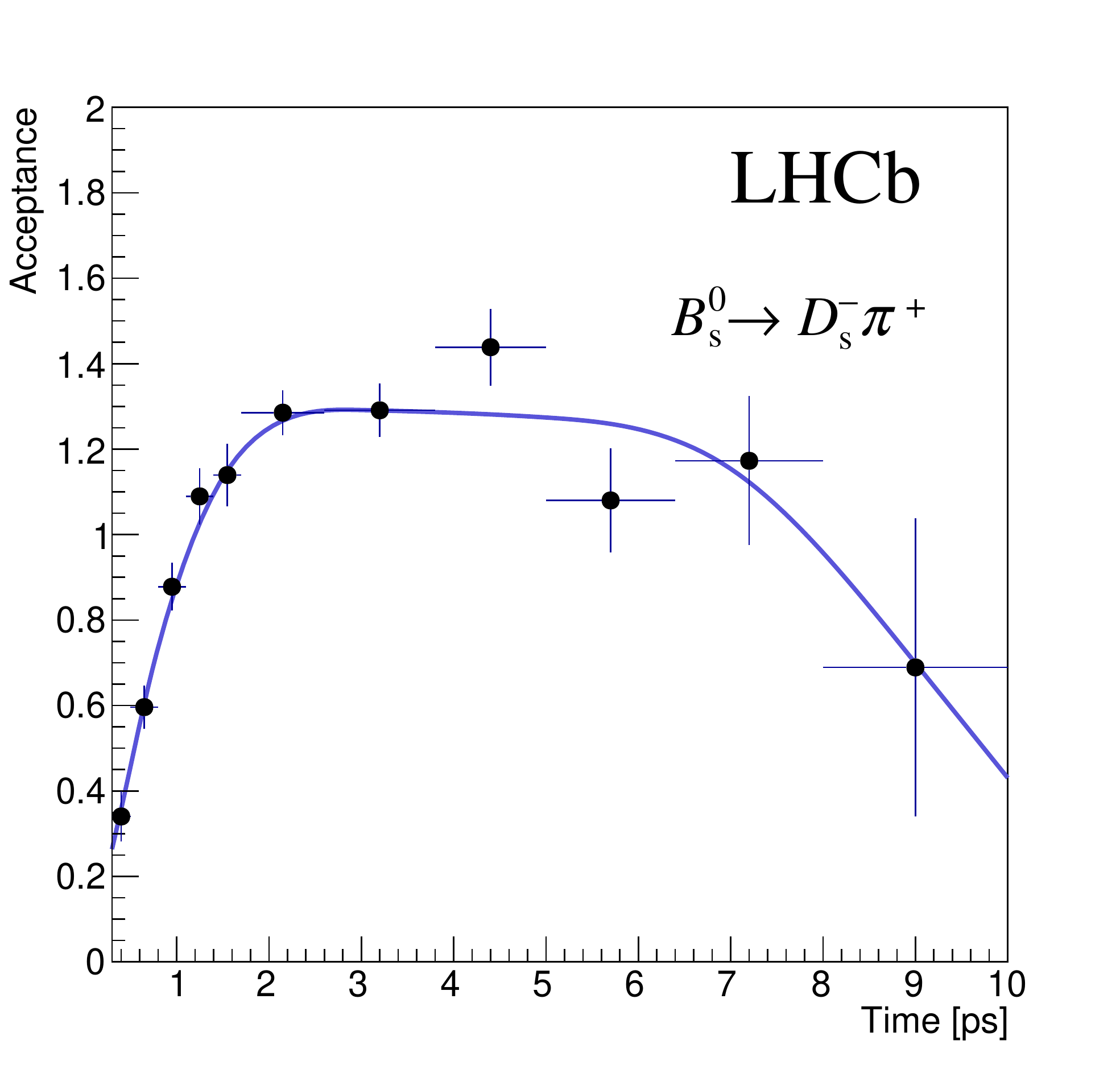}
\includegraphics[width=0.45\textwidth]{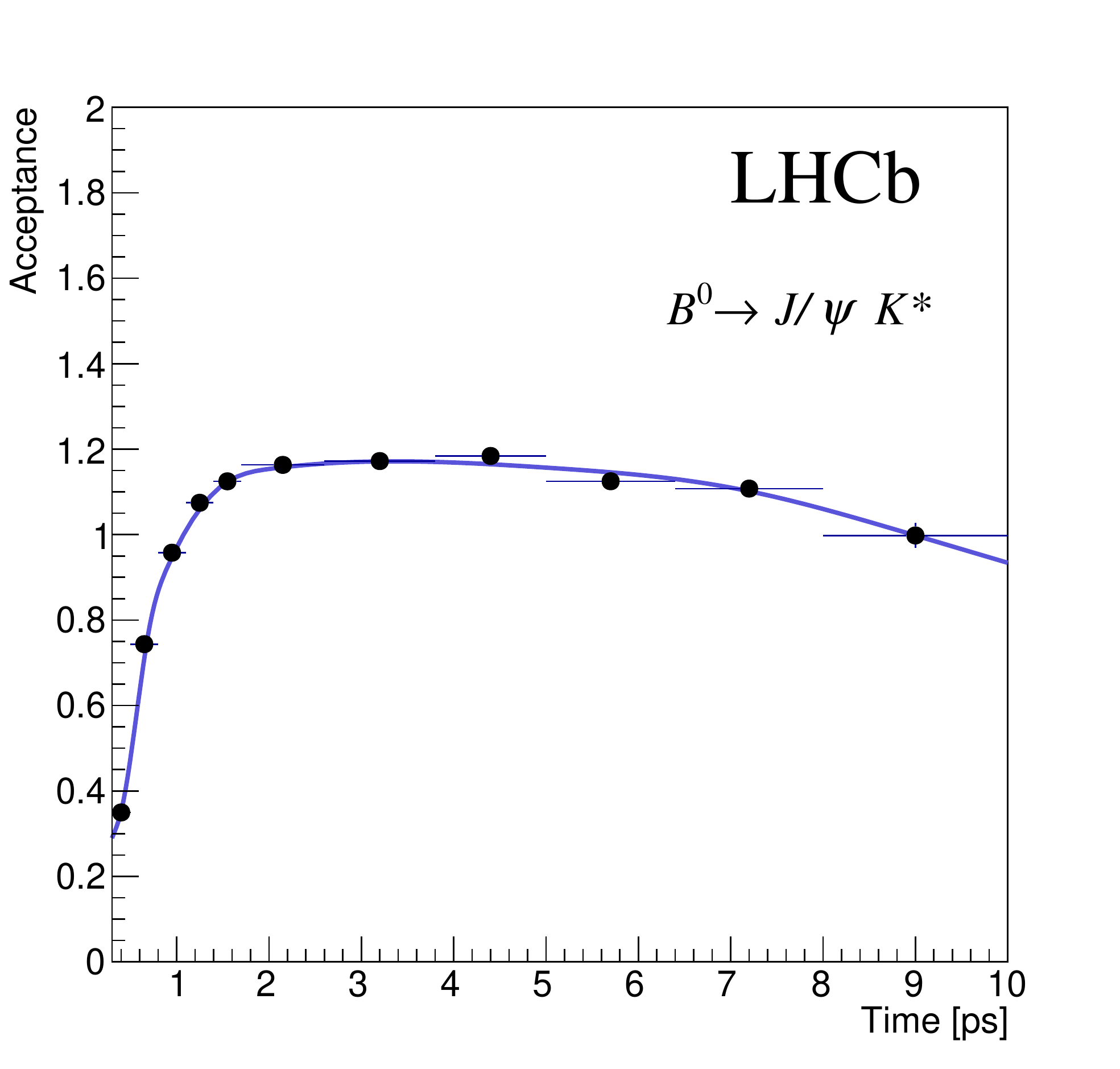}
\caption{
  \small Decay-time acceptances calculated from (left) $\Bs\to\Dsm\pip$ decays to match Run 1 data and 
  (right) $\Bd\to\jpsi\Kstarz$ decays to match Run 2 data.
  Superimposed is a parameterisation using cubic splines..
}
\label{fig:TA}
\end{figure}

Cubic splines are used to model the acceptance as a function of decay time in the PDF. The PDF can then be computed analytically with the
inclusion of the decay-time acceptance following Ref.~\cite{GRnote}.
%The use of cubic splines allows the acceptance to be constructed as
%
%\begin{align}
%\epsilon(t) = \sum_{i=0}^6 c_i b_i(t),
%\end{align}
%
%where $c_i$ are coefficients and $b_i(t)$ are third-order polynomial functions defined over
%distinct time ranges. The cubic spline acceptance description is chosen to have 9 subintervals, thus using 10 knots. 
%These are chosen to be at: $0.31,\, 0.4,\, 0.7,\, 0.9,\, 1.5,\, 2.0,\, 3.0,\, 5.0,\, 7.0,\, 9.0$ ps. 
%With these knots,
%the coefficients are calculated as described in Ref.~\cite{GRnote}.
Example decay-time acceptances are shown for the case of the $\Bs\to\Dsm\pip$ and
$\Bd\to\jpsi\Kstarz$ decays in Fig.~\ref{fig:TA}.

To simplify the measurement of the triple-product asymmetries, the decay-time acceptance is not applied in the fit to determine the triple-product asymmetries. The time acceptance correction has an impact on the asymmetry of 0.3$\%$ and is treated as a source of systematic uncertainty, as further described in Sec.~\ref{sec:systTP}.

\section{Flavour tagging}
\label{sec:flavour}

To obtain sensitivity to \phisPP, the flavour of the \Bs meson at production must be determined.
%The terms in the differential decay rate with the largest sensitivity to \phisPP
%require the identification (tagging) of the flavour at production.
At \lhcb, tagging is achieved through the use of various algorithms
described in Refs.\cite{LHCB-PAPER-2011-027,LHCB-PAPER-2015-056}. 
With these algorithms, the flavour-tagging power, defined as $\epsilon_{\rm tag} \mathcal{D}^2$ can
 be evaluated. Here, $\epsilon_{\rm tag}$ is the
flavour-tagging efficiency defined as the fraction of candidates with a flavour tag with respect to the total,
and $\mathcal{D}\equiv (1-2\omega)$ is the dilution, where $\omega$ is the average fraction of candidates with an incorrect flavour
assignment.
This analysis uses opposite-side (OS) and same-side kaon (SSK) flavour taggers.

The OS flavour-tagging algorithm~\cite{LHCB-PAPER-2011-027} makes use of the \bquarkbar (\bquark) hadron 
produced in association with the signal \bquark (\bquarkbar) hadron. In this analysis, the predicted 
probability of an incorrect flavour assignment, $\omega$, is determined for each candidate by a neural network that is calibrated using $\Bu \to \jpsi \Kp$, 
$\Bu\to\Dzb\pip$, $\Bd\to\jpsi\Kstarz$, $\Bd\to\Dstarm\mup\neum$, and $\Bs \to \Dsm\pip$ data as control modes.
Details of the calibration procedure can be
found in Ref.~\cite{LHCb-PAPER-2014-059}. 

When a signal \Bs meson is formed, there is an associated \squarkbar quark formed in the first branches of the fragmentation that about 50\,\% of the time
forms a charged kaon, which is likely to originate close to the \Bs meson production point. 
The kaon charge therefore allows for the identification of the flavour of the signal \Bs meson.
This principle is exploited by the SSK flavour-tagging algorithm~\cite{LHCB-PAPER-2015-056}, which is calibrated with the $\Bs \to \Dsm\pip$ decay mode.
A neural network is used to select fragmentation particles, 
improving the flavour-tagging power quoted in the previous decay-time-dependent measurement~\cite{LHCb-PAPER-2014-026}.

Table~\ref{tab:TagPow} shows the tagging power for the candidates tagged by only one of the algorithms and those tagged by both.
Uncertainties due to the calibration of the flavour tagging algorithms are applied as Gaussian
constraints in the decay-time-dependent fit. The
initial flavour of the \Bs meson established from flavour tagging is accounted for during fitting.

\newcommand{\tabincell}[2]{\begin{tabular}{@{}#1@{}}#2\end{tabular}}

%%%%%%%%%%%%%%%%%%%%%%%%%%%%%%%%%
\def\TexPTag {tex/ss_nnetk_tagpower_partex}
\begin{table}[t]%\footnotesize
  \renewcommand\arraystretch{1.2}
  \begin{center}
    \caption{ \normalsize Tagging performance of the opposite-side (OS) and same-side kaon (SSK)
    flavour taggers for the \BsPP decay.}
  \label{tab:TagPow}
  \begin{tabular}{l|ccc}
    %\hline
    Category     %& Fraction (\%) 
    & $\varepsilon$ (\%)  & $\mathcal{D}^{2}$   & $\varepsilon\mathcal{D}^{2}$ (\%) \\
    \hline                                                           
    OS-only 
    %&\input{\TexPTag/p_osofra.tex} 
    &$12.5$ 
    &$0.10$ 
    &$1.24 \pm 0.10$ \\
    SSK-only 
    %&\input{\TexPTag/p_ssofra.tex} 
    &$41.0$ 
    &$0.04$ 
    &$1.74\pm0.36$ \\
    OS\&SSK 
    %&\input{\TexPTag/p_oasfra.tex} 
    &$23.3$ 
    &$0.12$ 
    &$2.76\pm0.20$ \\
    \hline
    Total 
    %&100 
    &$76.8$ 
    &$0.08$ 
    &$5.74\pm0.43$ \\
    %\hline
  \end{tabular}
  \end{center}
\end{table}
%%%%%%%%%%%%%%%%%%%%%%%%%%%%%%%%%

\section{Decay-time-dependent measurement}
%\label{sec:LL}
%
%Separate fits are performed in order to determine the triple-product asymmetries and the decay time dependent fit parameters.
%These are described separately in the following two sections.

\subsection{Likelihood fit}
\label{sec:LL_TD}

The fit parameters in the polarisation-independent fit are the \CP violation parameters, $\phisPP$ and $|\lambda|$, the squared amplitudes, $|A_0|^2$, $|A_\perp|^2$, $|A_S|^2$, and $|A_{SS}|^2$, 
and the strong phases, $\delta_\perp$, $\delta_\parallel$, $\delta_0$, $\delta_S$, and $\delta_{SS}$, as
defined in Sec.~\ref{sec:modelTD}. The $P$-wave amplitudes are defined such that
$|A_0|^2+|A_\perp|^2+|A_\parallel|^2=1$, hence only two of the three amplitudes are free parameters.
This normalisation effectively means the $S$ and $SS$ components are measured relative to the $P$-wave.
The polarisation-dependent fit allows for a perpendicular, parallel and longitudinal component of \phisPP and $|\lambda|$.

The measurement of the parameters of interest is performed through an unbinned negative log likelihood minimisation.
The log-likelihood, $\mathcal{L}$, of each candidate is weighted using the \sPlot method~\cite{Pivk:2004ty,Xie:2009rka}, to remove partly reconstructed and combinatorial background.
The negative log-likelihood then takes the form
\begin{align}
-\ln \mathcal{L} = -\alpha \sum_{e\in{\rm candidates}} W_e \ln (S^e_{\rm TD}),
\end{align}
where $W_e$ are the signal \sPlot weights calculated using the four-kaon invariant mass as the discriminating variable.
The correlations between the angular and decay-time variables used in the fit with the four-kaon mass are small enough for this technique to be appropriate. The factor $\alpha=\sum_eW_e/\sum_eW_e^2$ accounts for 
the \sPlot weights in the determination of the statistical uncertainties.
The parameter $S^e_{\rm TD}$ is the differential decay rate of Eq.~\ref{eq:pdf}, modified to the effects of decay-time and
angular acceptance, in addition to the probability of an incorrect flavour tag. Explicitly, this can be written as
\begin{align}
S^e_{\rm TD} = \frac{\sum_is^e_i(t_e)f_i(\Omega_e)\epsilon(t_e)}{\sum_k\zeta_k\int s_k(t)f_k(\Omega)\epsilon(t) \deriv t \, \deriv\Omega},
\end{align}
where $\zeta_k$ are the normalisation integrals used to describe the angular acceptance described in Sec.~\ref{sec:AA} and
\begin{align}
s^e_i(t)&=N_ie^{-\Gs t_e} \left[ c_iq_e(1-2\omega_e) \cos(\dms t_e) + d_iq_e(1-2\omega_e)\sin(\dms t_e) 
\right.
\nonumber\\ &\qquad\left.
+  a_i \cosh\left(\frac{1}{2}\DGs t_e\right)  + b_i \sinh\left(\frac{1}{2}\DGs t_e\right)\right] \otimes R(\sigma^{\rm cal}_e,t_e).
\label{eq:explicit}
\end{align}
The calibrated probability of an incorrect flavour assignment is given by $\omega_e$, $R$ denotes the Gaussian time-resolution function,
and the $\otimes$ denotes a convolution operation.
In Eq.~\ref{eq:explicit}, $q_e=1$ $(-1)$ for a \Bs (\Bsb) meson at $t=0$ or $q_e=0$ where no flavour-tagging information is assigned.
The data samples corresponding to the different years of data taking are assigned independent signal weights, decay-time and angular acceptances,
and separate Gaussian constraints are applied to the decay-time resolution parameters, as defined in Sec.~\ref{sec:DTR}.
The  \Bs-\Bsb oscillation frequency is constrained to the value measured by \lhcb 
of $\dms=17.768\pm0.023\stat\pm0.006\syst\invps$~\cite{LHCb-PAPER-2013-006}, with the assumption
that the systematic uncertainties are uncorrelated with those of the current measurement. The values of the decay
width and decay-width difference are constrained to the current best known values of $\Gamma_s=0.6646 \pm 0.0020$\invps and $\Delta \Gamma_s = 0.086 \pm 0.006$\invps~\cite{HFLAV16}.

Correction factors must be applied to each of the $S$-wave and double $S$-wave interference terms in the differential decay width.
These factors modulate the sizes of the contributions of the interference terms in the angular PDF due to the 
different line-shapes of kaon pairs originating from spin-1 and spin-0 configurations. 
This takes the form of a multiplicative factor for each time a $S$-wave pair of kaons
interferes with a $P$-wave pair.
Their $K^+K^-$ invariant-mass parameterisations are denoted by $g(m_{\Kp\Km})$ and $h(m_{\Kp\Km})$, respectively. The $P$-wave
configuration is described by a Breit--Wigner function and the $S$-wave configuration is assumed to be uniform.  
The correction factors, denoted by $C_{SP}$, are defined in Ref.~\cite{LHCb-PAPER-2014-059}
\begin{align}
%C_{SP}e^{i\theta_{SP}} = \int^{m_{\rm h}}_{m_{\rm l}} g^*(m_{\Kp\Km}) h(m_{\Kp\Km}) \deriv m_{\Kp\Km},
C_{SP} = \int^{m_{\rm h}}_{m_{\rm l}} g^*(m_{\Kp\Km}) h(m_{\Kp\Km}) \deriv m_{\Kp\Km},
\end{align}
%
%where $m_{\rm h}$ and $m_{\rm l}$ are the upper and lower edges of a given $m_{\Kp\Km}$ bin and $\theta_{SP}$ is absorbed in the measurements of $\delta_S - \delta_{\perp}$.
where $m_{\rm h}$ and $m_{\rm l}$ are the upper and lower edges of the $m_{\Kp\Km}$ window
and the phase of $C_{SP}$ is absorbed in the measurements of $\delta_S - \delta_{\perp}$.
The factor $|C_{SP}|$, is calculated to be 0.36.  In order to determine systematic
uncertainties due to the model dependence of the S-wave, $C_{SP}$ factors are recalculated based
on the $S$-wave originating from an $f_0(980)$ resonance and incorporating the effects of the $m_{K^+K^-}$ resolution.
These alternative assumptions on the $P$-wave and $S$-wave lineshapes  yield a $|C_{SP}|$ value of 0.34, which is found to
have a negligible effect on the parameter estimation.

%for the single $m_{\Kp\Km}$ bin used in this analysis.

\subsection{Results}
\label{sec:resTD}

The resulting parameters
are given in Table~\ref{tab:nominal}. A polarisation-independent fit is performed to calculate
values for $\phisPP$ and $|\lambda|$. A negligible fraction of $S$-wave and double $S$-wave is observed.

\begin{table}[b]\centering
\caption{\small \label{tab:nominal} Results of the decay-time-dependent, polarisation-independent fit
  for the \CP-violation fit.
  Uncertainties shown do not include systematic contributions.}
\begin{tabular}{@{}lc}
Parameter &  Fit result \\ \hline
\phisPP [rad]     & $-0.073      \pm 0.115  ~ ~    $ \\
$|\lambda|$  & $0.99      \pm 0.05      $ \\\hline
%$\phipa$ [rad]        & $0.014      \pm 0.055      $\\
%$\phipe$ [rad]        & $0.044      \pm 0.059      $ \\\hline
\az          & $0.381    \pm 0.007    $ \\
\ape         & $0.290    \pm 0.008    $ \\\hline
%\dpe  [rad]       & $2.818      \pm 0.178      $ \\
%\dpa  [rad]       & $2.559      \pm 0.045      $ \\
\dpe  [rad]       & $2.82      \pm 0.18      $ \\
\dpa  [rad]       & $2.56      \pm 0.05      $ \\
\end{tabular}
\end{table}

In addition, the \CP-violating phases are also determined
in a polarisation-dependent manner. Due to limited size of data samples, the phases $\phipa$ and $\phipe$ are measured under the assumptions that the longitudinal
weak phase is \CP-conserving and that there is no direct \CP violation. In addition, all $S$-wave and double $S$-wave components of the fit are set to zero. The results of the polarisation dependent fit are shown in Table~\ref{tab:poldep}. The results for \az, \ape, \dpe and \dpa are not shown but are in agreement with the results reported in Table~\ref{tab:nominal}.

\begin{table}[t]\centering
\caption{\small \label{tab:poldep} Results of the polarisation-dependent fit
  for the \CP violation fit.
  Uncertainties shown do not include systematic contributions. }
\begin{tabular}{@{}lc}
Parameter &  Fit result \\ \hline
%\phisPP [rad]     & $-0.073      \pm 0.115  ~ ~    $ \\
%$|\lambda|$  & $0.99      \pm 0.05      $ \\\hline
$\phipa$ [rad]        & $0.014      \pm 0.055      $\\
$\phipe$ [rad]        & $0.044      \pm 0.059      $ \\
%\az          & $0.381    \pm 0.007    $ \\
%\ape         & $0.290    \pm 0.008    $ \\\hline
%\dpe  [rad]       & $2.818      \pm 0.178      $ \\
%\dpa  [rad]       & $2.559      \pm 0.045      $ \\
%\dpe  [rad]       & $2.82      \pm 0.18      $ \\
%\dpa  [rad]       & $2.56      \pm 0.05      $ \\
\end{tabular}
\end{table}

The correlation matrices for the two fit configurations are provided in Appendix~\ref{app:corr}. 
Correlations with such decay-time-dependent measurements depend on the central values
of the parameters. No large correlation is expected between the \CP-violating
parameters when the central values are consistent with \CP conservation. 
The largest correlations are found to be between the 
different decay amplitudes.
%The observed correlations have been verified with simulated data sets.
Cross-checks are performed on simulated data sets generated with the same yield as observed in data,
and with the same physics parameters, to establish that the generated values are recovered without biases.

\begin{figure}[t]
\begin{center}
\includegraphics[width=0.48\textwidth]{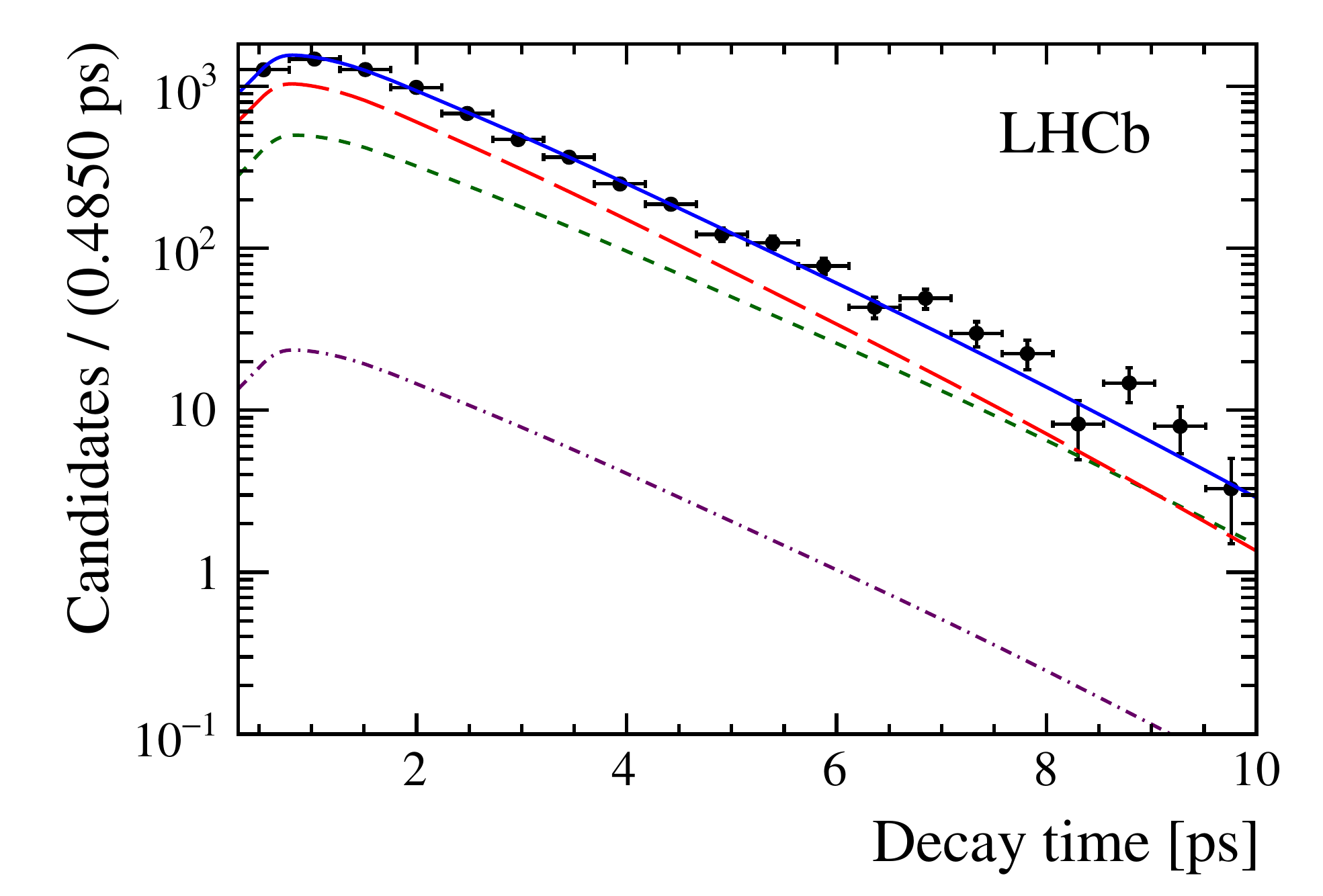}
\includegraphics[width=0.48\textwidth]{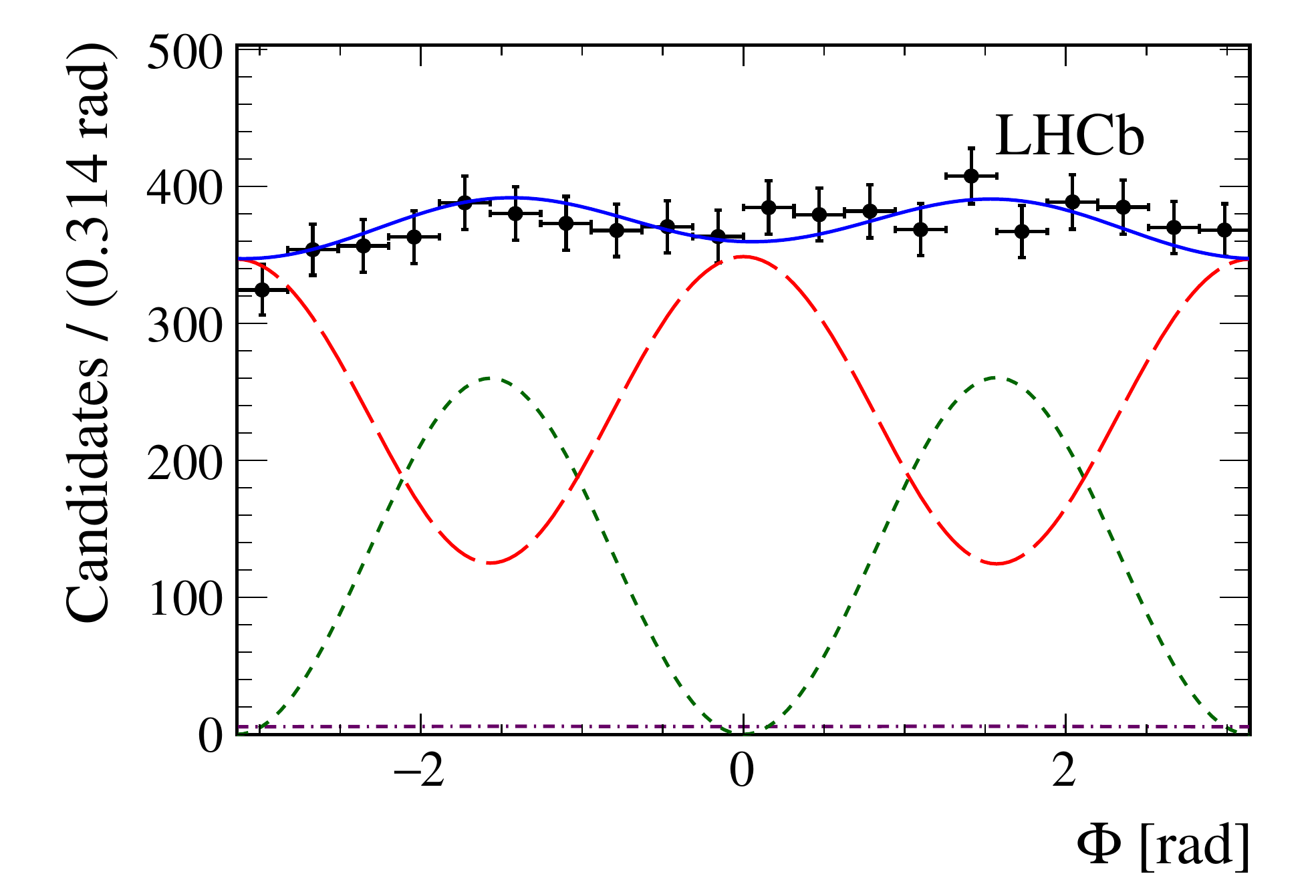}\\
\includegraphics[width=0.48\textwidth]{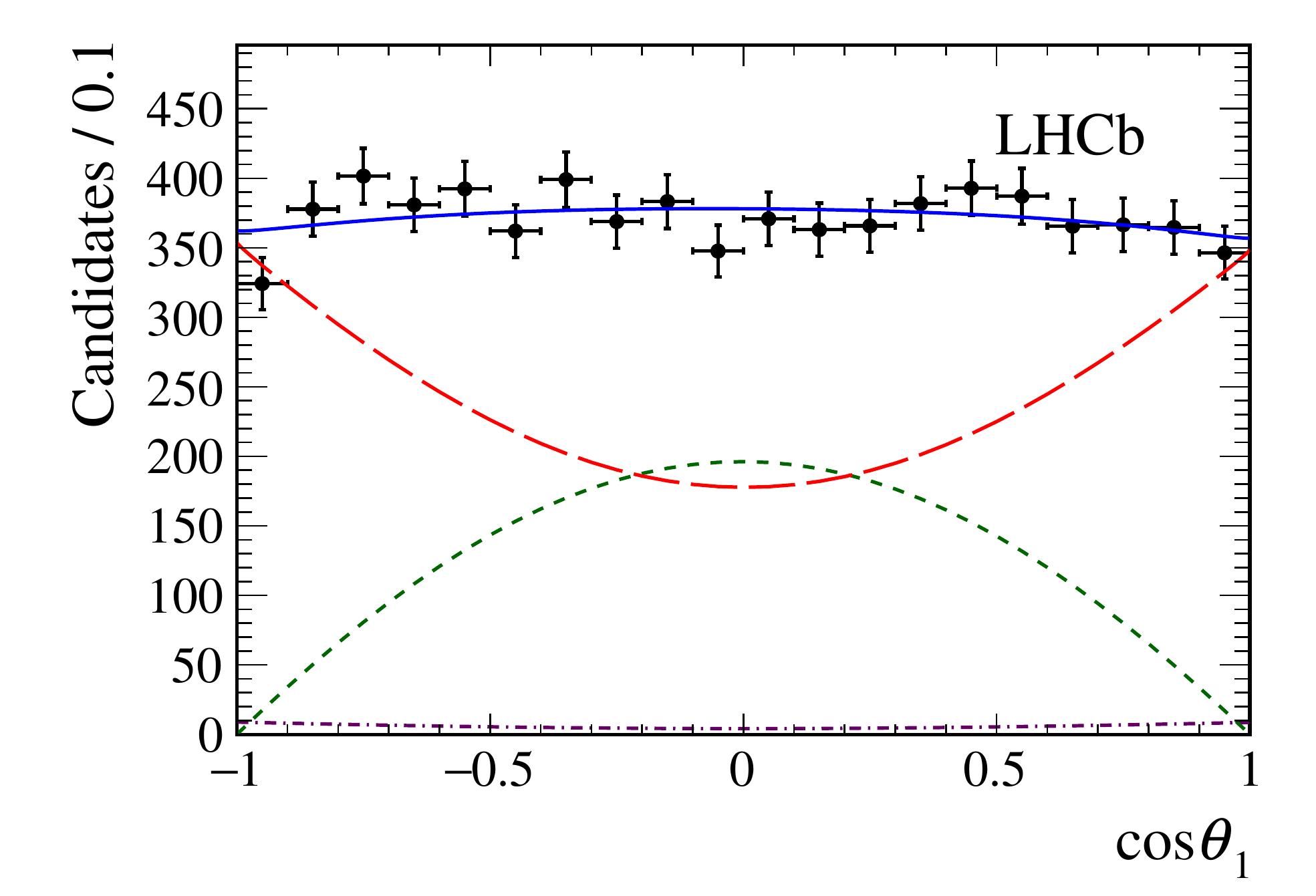}
\includegraphics[width=0.48\textwidth]{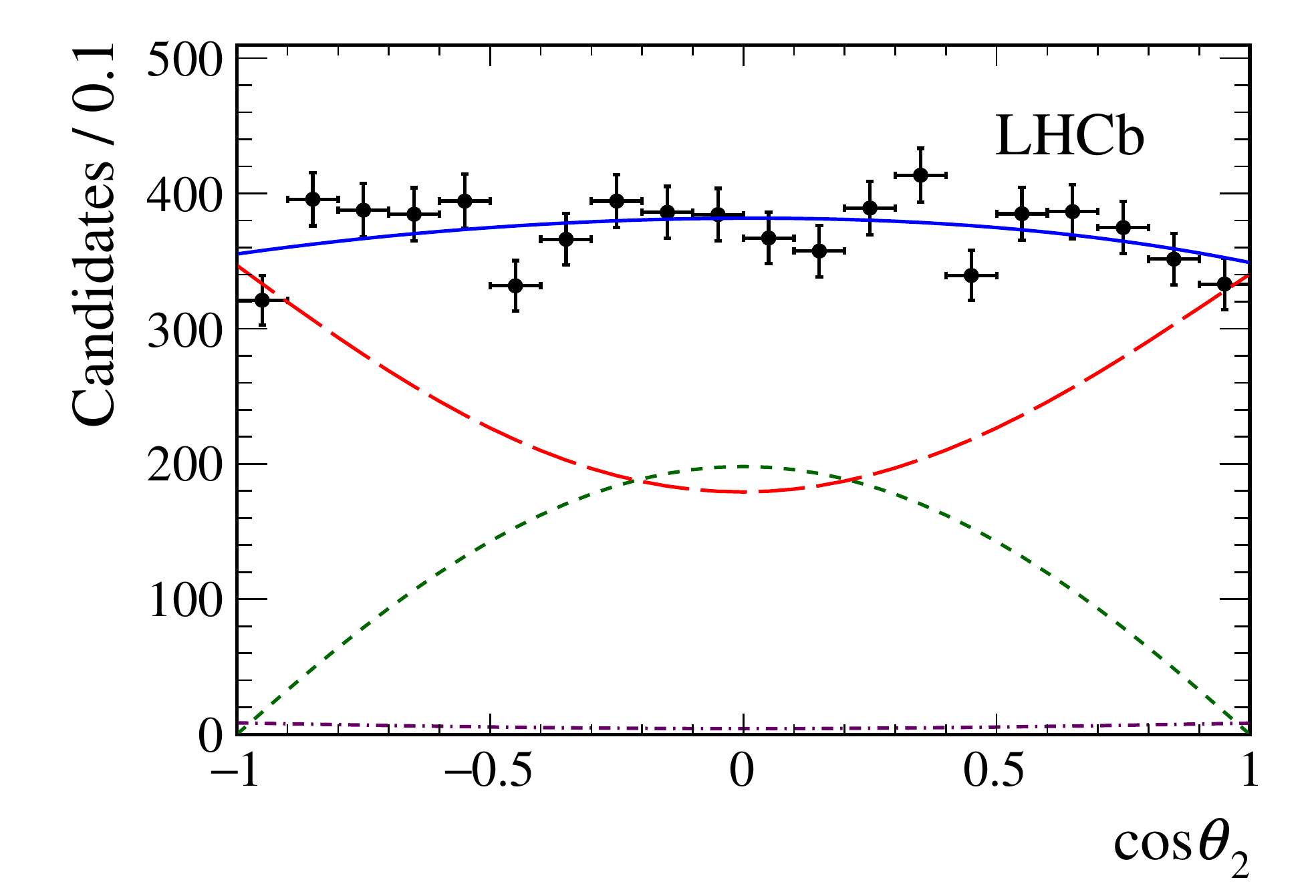}
\caption{One-dimensional projections of the $\Bs \to \phi\phi$ fit for (top-left) decay time with binned acceptance,
  (top-right) helicity angle $\Phi$ and (bottom-left and bottom-right) cosine
    of the helicity angles $\theta_1$ and $\theta_2$.
       The background-subtracted data are marked as black points, while the blue solid lines represent the projections of the fit.
          The \CP-even $P$-wave, the \CP-odd $P$-wave and the combined $S$-wave and double $S$-wave
	     components are shown by the red long dashed, green short dashed and purple dot-dashed lines, respectively.
             Fitted components are plotted taking into account efficiencies in the time and angular observables.}
\label{fig:proj}
\end{center}
\end{figure}

Figure~\ref{fig:proj} shows the distributions of the \Bs decay time and the three
helicity angles.
Superimposed are the projections of the fit result. The projections 
%are weighted to yield the signal distribution
%and 
include corrections for acceptance effects. Pseudoexperiments were used to confirm that the deviation of the data around $\cos\theta_2 = \pm 0.5$ from the resulting distribution of the fit is compatible with a statistical fluctuation.

\subsection{Systematic uncertainties}
\label{sec:syst}

Various sources of systematic uncertainty are considered in addition to those
applied as Gaussian constraints in the fit.
These arise from the angular and decay-time acceptances,
the mass model used to describe the mass distribution, the determination of the time resolution calibration, and 
the fit bias. A summary of the systematic uncertainties is given in Table~\ref{tab:syst_sum}. 

An uncertainty due to the angular acceptance arises from the choice of weighting scheme
described in Sec.~\ref{sec:acc}. 
This is accounted for by performing multiple alternative weighting schemes 
for the weighting procedure, based on different kinematic variables in the decay. The largest variation
is then assigned as the uncertainty.
Further checks are performed to verify that the angular acceptance does not depend
on the way in which the event was triggered.

Two sources of systematic uncertainty are considered concerning the decay-time
acceptance. These are the statistical uncertainty from the spline coefficients, and also
the residual disagreement between the weighted control mode and the signal decay acceptances (see Sec.~\ref{sec:DTA}).
The former is evaluated through fitting the signal data set with 30 different spline functions,
whose coefficients are varied according to the corresponding uncertainties. This study is performed with
the three different choices of the knot points.
The RMS of the fitted parameters is then assigned as the uncertainty. 
The residual disagreement between the control mode and the signal 
mode is accounted for with a simulation-based correction. 
Simplified simulation is used with the corrected acceptance and then fitted with the nominal acceptance.
This process is repeated and the resulting bias on the fitted parameters
is used as an estimate of the systematic uncertainty.

The uncertainty on the mass model is found by refitting the data with various alternative
signal models, consisting of the sum of two Crystal Ball models, the sum of a double-sided
Crystal Ball and a Gaussian model. In addition, a Chebyshev polynomial is used to describe the
combinatorial background.
The signal weights are recalculated and the largest deviation from the nominal fit
results is used as the corresponding uncertainty.

Fit biases can arise in maximum-likelihood fits where the number of candidates is small compared to the number of free parameters.
The effect of such a bias is taken as a systematic uncertainty which is evaluated by generating 
and fitting simulated data sets and taking the resulting bias as the uncertainty.

The uncertainties of the effective flavour-tagging power are included in the statistical uncertainty through
Gaussian constraints on the calibration parameters, and amount to 10\,\% of the statistical
uncertainty on the \CP-violating phases.

\begin{table}[t]
\centering
\caption{\small Summary of systematic uncertainties (in units of $10^{-2}$) for parameters of interest in the decay-time-dependent measurement.}
          {
	\begin{tabular}{c|ccccc|c}
          Parameter 	 & Mass  	        & Angular 	& Decay-time 	& Time & Fit  & Total \\ 
           & model & acceptance & acceptance & resolution & bias &  \\ \hline
          $|A_0|^2$      & 0.4  & 1.1  	&  0.1  & -     & 0.2& 1.2\\
          $|A_\perp|^2$  & -    & 0.5  	&  -    &  -    & 0.1& 0.5\\
          $\dpa$ [rad]   & 2.7  & 0.2 	&  0.5  & 0.1   & 1.7& 3.3\\
          $\dpe$ [rad]   & 3.8  & 0.3  	&  0.8  & 1.4   & 6.0& 7.3\\
          \phisPP [rad]  & 1.2  & 0.5  	&  0.6  & 2.0   & 1.1& 2.7\\
          $\lambda$      & 0.5  & 0.5  	&  0.2  & 0.3   & 0.9& 1.2\\\hline
          \phipa [rad]   & 0.2  & 0.2  	&  0.4  & 0.2   & 1.0& 1.1\\
          \phipe [rad]   & 1.4  & 0.3  	&  0.4  & 0.3   & 0.4& 1.9\\
        \end{tabular}
        }
\label{tab:syst_sum}
\end{table}

\section{Triple-product asymmetries}
\subsection{Likelihood}
\label{sec:LL_TPA}

To determine the triple-product asymmetries, the data sets are divided according to the
sign of the observables $U$ and $V$. Simultaneous likelihood fits to the four-kaon mass distributions are preformed
for the $U$ and $V$ variables separately.
%In order to determine the triple-product asymmetries, a separate likelihood fit is performed. 
%A simultaneous fit of separate data sets to the four-kaon invariant mass is performed,
%which are split according to the sign of $U$ and $V$ observables.
%Simultaneous mass fits are performed for the $U$ and $V$ observables separately.
The set of free parameters in the fits to determine the $U$ and $V$ observables consists of their total yields and
the asymmetries $A_{U(V)}$.
The mass model is the same as that described in Sec.~\ref{sec:selection}.
The total PDF, $D_{\rm TP}$, is then of the form
\begin{equation}
D_{\rm TP} = \sum_{i \in \{+,-\}} \left( f_i^S G^S(m_{\Kp\Km\Kp\Km}) + \sum_{k}f^k_i P^k(m_{\Kp\Km\Kp\Km}) \right),
\label{eq:newtotpdf}
\end{equation}
where $k$ indicates the sum over the background components with corresponding PDFs, $P^j$,
and $G^S$ is the signal PDF, as described in Sec.~\ref{sec:selection}.
The parameters $f^S_i$ found in Eq.~\ref{eq:newtotpdf} are related to the asymmetry, $A_{U(V)}^S$, through
\begin{eqnarray}
  f_{U(V),+}^{S} = \frac{1}{2}(1 + A^S_{U(V)}), \\
  f_{U(V),-}^{S} = \frac{1}{2}(1-A^S_{U(V)}),
\label{eq:asymmetries1}
\end{eqnarray}
where $S$ denotes the signal component of the four-kaon mass fit, as described in Sec.~\ref{sec:selection}. 
Peaking backgrounds are assumed to be symmetric in $U$ and $V$.

\subsection{Results}
\label{sec:resTP}

The triple-product asymmetries found from the simultaneous fit described in Sec.~\ref{sec:LL_TPA} are measured separately for the 2015 and 2016 data.
The results are combined by performing likelihood scans of the asymmetry
parameters and summing the two years. This gives
\begin{center}
\begin{tabular}{l@{$~=~$}r@{$\,\,\pm\,$}l}
$A_U$ & $-$0.003 & 0.015 ,  \\
$A_V$ & $-$0.012 & 0.015 ,
\end{tabular}
\end{center}
where the uncertainties are statistical only.

\subsection{Systematic uncertainties}
\label{sec:systTP}

As for the case of the decay-time-dependent fit, significant contributions to
the systematic uncertainty arise from the decay-time and angular acceptances.
Minor uncertainties also result from the knowledge of the mass model of the signal and the composition of peaking backgrounds.

The effect of the decay-time acceptance is determined through the generation of
simulated samples including the decay-time acceptance
and fitted with the method described in Sec.~\ref{sec:LL_TPA}.
The resulting deviation from the nominal fit results is used to assign a systematic uncertainty. % and is determined to be 0.3\,\%.
The effect of the angular acceptance is evaluated by generating simulated data sets with and without
the inclusion of the angular acceptance. 
The difference between the nominal fit results and the results obtained using the simulated samples including the decay-time acceptance is then used as a systematic uncertainty.

Uncertainties related to the mass model are evaluated
using a similar approach to that described in Sec.~\ref{sec:syst}.
%and are determined to be 0.1\,\%.
Additional uncertainties arise from the assumption that the peaking background is symmetric in $U$ and $V$.
The deviation observed without this assumption is then added as a systematic uncertainty. 
%These are determined to be 0.1\,\%. 
Similarly, the assumption that the combinatorial background has no asymmetry yields an identical uncertainty.
The systematic uncertainties are summarised in Table~\ref{tab:TPA_systtable}.

\begin{table}[tb]
\centering
\caption{Summary of systematic uncertainties on $A_U$ and $A_V$.}
\begin{tabular}{l|c} 
Source &  Uncertainty  \\ \hline
Time acceptance & 0.003 \\
Angular acceptance & 0.003 \\
Mass model & 0.001 \\
Combinatorial background & 0.001 \\
Peaking background & 0.001 \\ \hline
Total & 0.005  \\ 
\end{tabular}
\label{tab:TPA_systtable}
\end{table}

\subsection{Combination of Run 1 and Run 2 results}
The Run 2 (2015--2016) values for the triple product asymmetries are
\begin{center}
\begin{tabular}{l @{$~=~$}l@{$\,\,\pm\,$}l@{\,(stat)
  $\pm\,$}l@{\,(syst)}lr}
$A_U$& $-$0.003 & 0.015 & 0.005&,  \\
$A_V$& $-$0.012 & 0.015 & 0.005&, \\
\end{tabular}
\end{center}
whilst the Run 1 (2011--2012) values from Ref.~\cite{LHCb-PAPER-2014-026} are
\begin{center}
\begin{tabular}{l @{$~=~$}l@{$\,\,\pm\,$}l@{\,(stat)
  $\pm\,$}l@{\,(syst)}lr}
$A_U$&$-$0.003 & 0.017 & 0.006&,  \\
$A_V$&$-$0.017 & 0.017 & 0.006&. \\
\end{tabular}
\end{center}

The Run 1 and Run 2 results are combined by calculating a weighted
average. In this procedure the decay-time and angular acceptance systematic uncertainties
and peaking backgrounds are assumed to be fully correlated.
All other systematic uncertainties are assumed to be
uncorrelated. This gives  a final result of
\begin{center}
\begin{tabular}{l@{$~=~$}r@{$\,\,\pm\,$}r@{$\,\,\pm\,$}l}
$A_U$ & $-$0.003 & 0.011\stat & 0.004\syst,  \\
$A_V$ & $-$0.014 & 0.011\stat & 0.004\syst.
\end{tabular}
\end{center}

The Run 1 and Run 2 results are compatible with each other, and the asymmetries are consistent with zero.
No evidence for \CP
violation is found.

\section{\protect\boldmath Search for the $B^{0} \rightarrow \phi \phi$ decay}
\label{sec:bd}

The selection criteria for the $B^0\rightarrow\phi \phi$ mode are based on the \BsPP selection, with some modifications.
The Punzi figure of merit~\cite{Punzi:2003bu} is used for the $\Bd\to\phi\phi$ search, resulting in a more stringent MLP requirement.
Furthermore, the uncertainty on the four-kaon mass is required to be less than 25\mevcc, corresponding to roughly $3\sigma$ separation between the $\Bs$ and $B^0$ mass peaks.
The \BsPP decay is used as normalisation decay mode. 
The signal PDF for the mass of the \Bd meson is assumed to be the same as that of the \Bs decay,
with the modification of the resolution according to a scaling factor, which is defined as
\begin{equation}
\alpha = \frac{m_{B^{0}}-4m_{K}}{m_{B^{0}_{s}}-4m_{K}}=0.974,
\end{equation}
where  $m_{K}$ is the known \Kp mass. 

\begin{figure}[t]
\begin{center}
\includegraphics[width=0.7\textwidth]{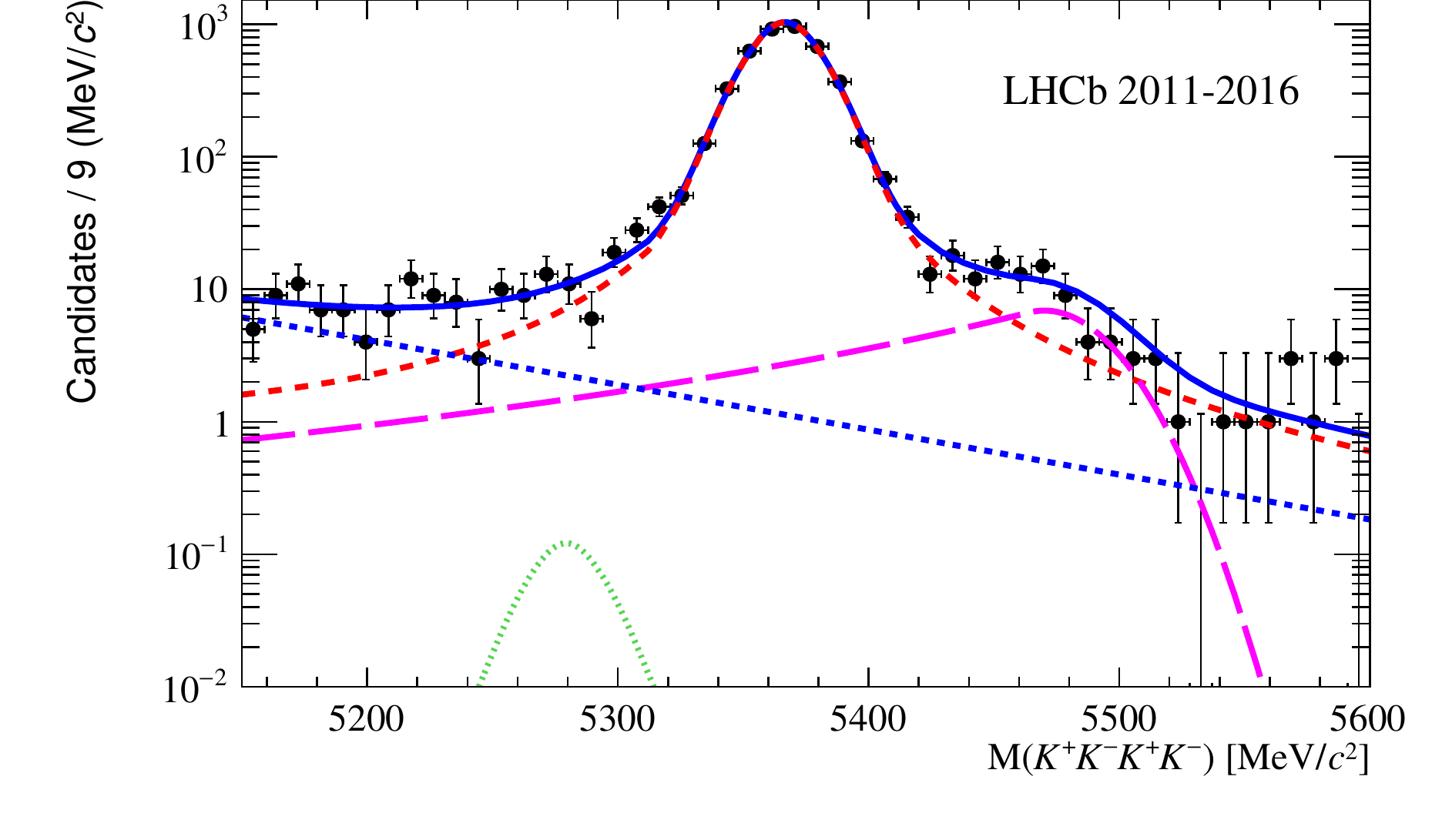}
\caption{Fit to the four-kaon invariant mass. The total PDF as described in the text is shown as a blue solid line, \BsPP as a red dashed line, $\Bd \rightarrow \phi \phi$ as a green dotted line, the $\Lb \rightarrow \phi p K$ contribution as a magenta long-dashed line and the combinatorial background as a blue short-dashed line. 
%Normalised residuals with respect to the total fit are given in the upper subpanel.
  }
\label{fig:bdfit}
\end{center}
\end{figure}

Figure~\ref{fig:bdfit} shows the fit to the full data set. The $\Lb \rightarrow \phi p K$ contribution is fixed to 
109 candidates, following the same method described in Sec.~\ref{sec:selection}. The fit returns
a yield of $4.9\pm9.2$ $\Bd\rightarrow \phi \phi$ decays.

The Confidence Levels (${\rm CL}_s$) method~\cite{CLs} is used to set a limit on the
$B^{0} \rightarrow \phi \phi$ branching fraction. 
%The test statistic used in the ${\rm CL}_s$ method is defined as 
%
%\begin{equation}
%t=-2\mathrm{log}\frac{\mathcal{L}(f_{s+b}(x))}{\mathcal{L}(f_{b}(x))%},
%\label{eq:t} 
%\end{equation}
%
%where $\mathcal{L}(f_{s+b}(x))$ is the likelihood of the signal plus %background PDF, $f_{s+b}$, and $\mathcal{L}(f_{b}(x))$ is the %likelihood calculated 
%from the background only PDF, $f_b$. The set of input parameters to %the PDF is denoted by $x$.
A total of 10,000 pseudoexperiments are used to calculate each point of the scan.
\begin{figure}[t]
\begin{center}
\includegraphics[width=0.7\textwidth]{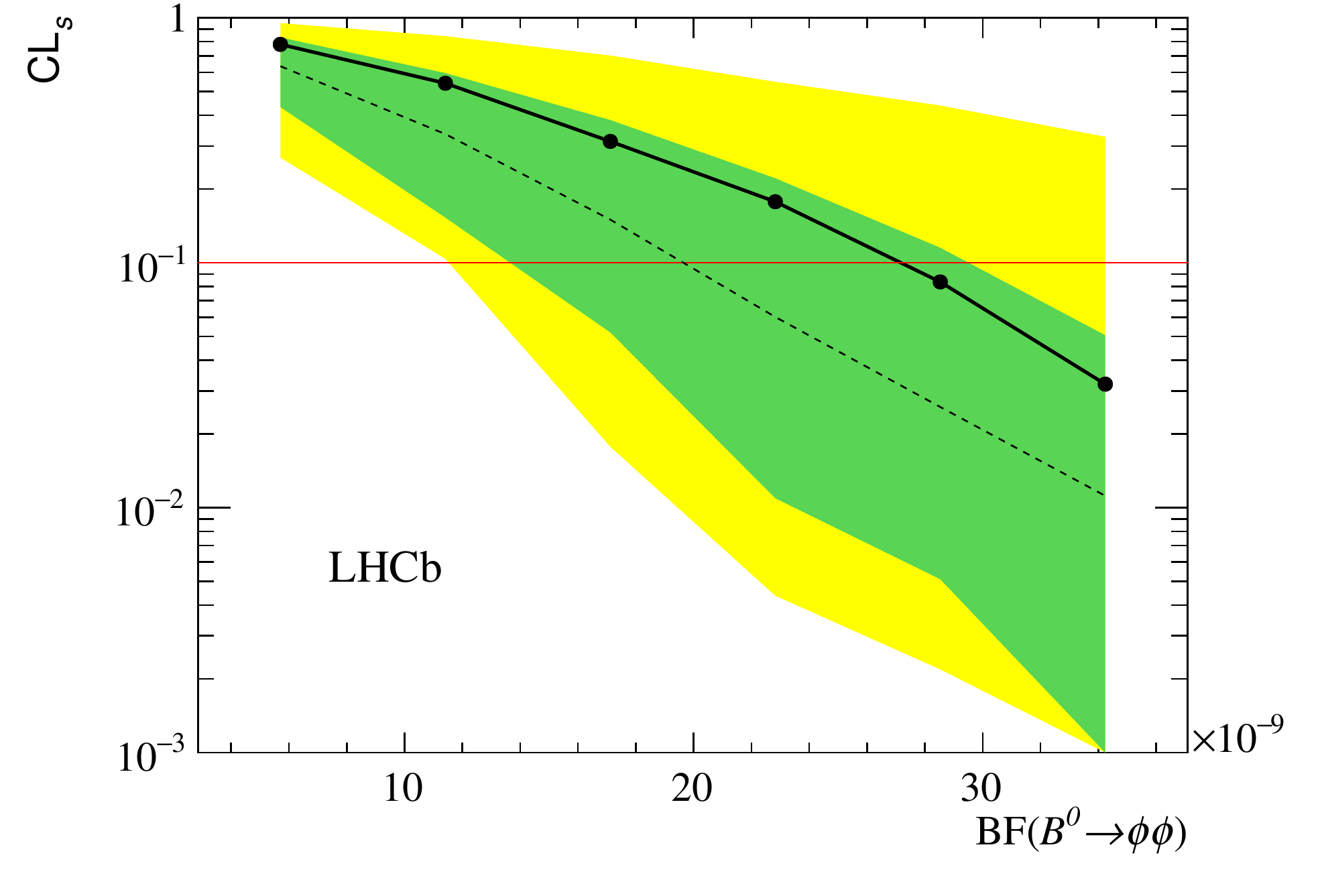}
  \caption{Results of the ${\rm CL}_s$ scan as a function of the $\Bd\rightarrow\phi\phi$ yield. The solid black line shows the observed ${\rm CL}_s$ distribution,
  while the dotted black line indicates the expected distribution. The green (yellow) 
  band marks the $1\sigma$ ($2\sigma$) confidence region on the expected ${\rm CL}_s$. The $90\,\%$ CL limit is shown as a red line.}
\label{fig:CLplot}
\end{center}
\end{figure}
Figure~\ref{fig:CLplot} shows the results of the ${\rm CL}_s$ scan. At $90\,\%$ {\rm CL}, $N_{\Bd}<23.7$. 
These limits are converted to a branching fraction using
\begin{equation}
\mathcal{B}(B^{0}\rightarrow \phi \phi) = N_{B^{0}} \times \frac{\epsilon_{B^0\rightarrow \phi \phi}}{ \epsilon_{\BsPP}}\times \frac{\mathcal{B}(B_{s}^{0}\rightarrow \phi \phi) \times f_s/f_d}{N_{B_{s}^{0}\rightarrow \phi \phi}},
\label{eq:bf} 
\end{equation}
where $N_{\Bd}$ is the limit on the 
$B^{0} \rightarrow \phi \phi$ yield, and $N_{\BsPP}$ is 
the $B_{s}^{0}$ yield from the fit displayed in Fig.~\ref{fig:bdfit}. The relative reconstruction and selection efficiency 
of the $\BsPP$ and $B^0\rightarrow \phi \phi$ decays, $\epsilon_{B^0\rightarrow \phi \phi}/ \epsilon_{\BsPP}$,  is determined to be 0.986 using
simulation. 
The ratio of the fragmentation functions has been measured at 7 and 8 \tev to be $f_s/f_d = 0.259 \pm 0.015$ within the LHCb acceptance~\cite{fsfd}. The production fraction at
13 \tev has been shown to be consistent with that of the 7 and 8 \tev data~\cite{LHCb-PAPER-2017-001}. The $\mathcal{B}(\BsPP)=(1.84 \pm 0.05\stat\pm 0.07\syst\pm 0.11 (f_s/f_d) \pm 0.12\,(\rm norm)) \times 10^{-5}$ 
branching fraction is an external input taken from Ref.~\cite{LHCb-PAPER-2015-028}.
To set the limit, the uncertainties on the $\BsPP$ branching fraction are propagated to the limit, 
where the uncertainty on the $\BsPP$ branching fraction arising from $f_s/f_d$ is already
included in the uncertainty on the normalisation mode, $\Bd\rightarrow \phi \Kstar$.
The maximum value of $\mathcal{B}(\BsPP)$ including the systematic contribution is found to be 
$1.99 \times 10^{-5}$ and is used in Eq.~\ref{eq:bf}. This therefore translates to a limit of
\begin{align*}
  \mathcal{B}(B^{0}\rightarrow \phi \phi) < 2.7 \,(3.0) \times 10^{-8} \, \mathrm{at}\, 90\,\% \, (95\,\%) \,{\rm CL},
\label{eq:bffinal} 
\end{align*}
which supersedes the previous best limit.
%of $2.8\times10^{-8}$ at 90\,\%
%CL~\cite{LHCb-PAPER-2015-028}. The presence of a significant combinatorial background 
%prevents a lower limit from being achieved, even with a significant increase in data set size.

\section{Summary and conclusions}
\label{sec:summary}

Measurements of \CP violation in the \BsPP decay are presented, based on a sample of proton-proton collision data corresponding to an integrated luminosity of   5.0\invfb collected with the \lhcb detector. The \CP-violating phase, \phisPP, and \CP violation parameter, $|\lambda|$,
are determined in a helicity-independent manner to be
\begin{center}
\begin{tabular}{l@{$~=~$}r@{$\,\,\pm\,$}r@{$\,\,\pm\,$}l}
\phisPP & $-$0.073 & 0.115\stat & 0.027\syst$\rad,$ \\
$|\lambda|$ & 0.99 & 0.05\stat & 0.01\syst$\kern-0.25em .$
\end{tabular}
\end{center}
The \CP-violating phases are also measured in a polarisation-dependent manner,
with the assumption that the longitudinal weak phase is \CP-conserving ($\philo =0$) and that no direct
\CP violation is present ($|\lambda|=1$). The \CP phases corresponding to the parallel, \phipa, and perpendicular, \phipe,
polarisations are determined to be
\begin{center}
\begin{tabular}{l@{$~=~$}r@{$\,\,\pm\,$}r@{$\,\,\pm\,$}l}
\phipa & 0.014 & 0.055\stat & 0.011\syst$\rad,$ \\
\phipe & 0.044 & 0.059\stat & 0.019\syst$\rad.$
\end{tabular}
\end{center}

The results are in agreement with SM predictions~\cite{Bartsch:2008ps,Beneke:2006hg,PhysRevD.80.114026}.
%Uncertainties are slightly larger than the naive expectation from %the scaling of the previous result~\cite{LHCb-PAPER-2014-026}.
The uncertainties have been validated with simulation.
%and the differences with respect to the expectation from the previous results are exaggerated through rounding effects.
When compared with the \CP-violating phase measured in ${\Bs\to\jpsi\Kp\Km}$ and ${\Bs\to\jpsi\pip\pim}$
decays~\cite{LHCb-PAPER-2014-059}, these results show that no significant \CP violation is present either in \Bs-\Bsb mixing or in the $\bquarkbar\to\squarkbar\ssbar$
decay amplitude, though the increased precision of the measurement presented
in Ref.~\cite{LHCb-PAPER-2014-059} leads to more stringent constraints on \CP violation in \Bs-\Bsb mixing.

The polarisation amplitudes and strong phases are measured independently of polarisation to be
\begin{center}
\begin{tabular}{l@{$~=~$} d{3} @{$\,\,\pm\,$} l@{$\,\,\pm\,$}l}
$|A_0|^2$ &         0.381 & 0.007\stat & 0.012\syst$\kern-0.25em ,$         \\
$|A_\perp|^2$ &     0.290 & 0.008\stat & 0.007\syst$\kern-0.25em ,$         \\
$\delta_\perp$ &        2.818 & 0.178\stat & 0.073\syst$\rad ,$        \\
$\delta_\parallel$ &    2.559 & 0.045\stat & 0.033\syst$\rad .$         
\end{tabular}
\end{center}
The polarisation amplitudes and strong phases measured in the polarisation-dependent fit are in agreement with the results listed here. 
In addition, values of the polarisation amplitudes are found to agree well with previous 
measurements~\cite{LHCb-PAPER-2012-004,LHCb-PAPER-2013-007,Aaltonen:2011rs,LHCb-PAPER-2014-026} 
and with predictions from QCD factorisation~\cite{Beneke:2006hg, PhysRevD.80.114026}.

The most precise measurements to date of the triple-product asymmetries are 
determined from a separate time-integrated fit to be
\begin{center}
\begin{tabular}{l@{$~=~$}r@{$\,\,\pm\,$}r@{$\,\,\pm\,$}l}
$A_U$ & $-$0.003 & 0.011\stat & 0.004\syst$\kern-0.25em ,$  \\
$A_V$ & $-$0.014 & 0.011\stat & 0.004\syst$\kern-0.25em ,$
\end{tabular}
\end{center}
in agreement with previous measurements~\cite{LHCb-PAPER-2012-004,Aaltonen:2011rs,LHCb-PAPER-2014-026}. 
The measured values of the \CP-violating phase and triple-product asymmetries are consistent 
with the hypothesis of \CP conservation in $\bquarkbar \to \squarkbar \ssbar$ transitions.

In addition, the most stringent limit on the branching fraction of the $\Bd\to\phi\phi$ decay is presented and it is found to be
\begin{equation*}
  \mathcal{B}(B^{0}\rightarrow \phi \phi)< 2.7 \times 10^{-8} \,\mathrm{at}\,90\,\% \,{\rm CL}.
\end{equation*}

% Comment this in for paper drafts; do not include this in analysis note and conference reports
\section*{Acknowledgements}
%
% These Acknowledgements valid from 3-May-2019
%
\noindent We express our gratitude to our colleagues in the CERN
accelerator departments for the excellent performance of the LHC. We
thank the technical and administrative staff at the LHCb
institutes.
We acknowledge support from CERN and from the national agencies:
CAPES, CNPq, FAPERJ and FINEP (Brazil); 
MOST and NSFC (China); 
CNRS/IN2P3 (France); 
BMBF, DFG and MPG (Germany); 
INFN (Italy); 
NWO (Netherlands); 
MNiSW and NCN (Poland); 
MEN/IFA (Romania); 
MSHE (Russia); 
MinECo (Spain); 
SNSF and SER (Switzerland); 
NASU (Ukraine); 
STFC (United Kingdom); 
DOE NP and NSF (USA).
We acknowledge the computing resources that are provided by CERN, IN2P3
(France), KIT and DESY (Germany), INFN (Italy), SURF (Netherlands),
PIC (Spain), GridPP (United Kingdom), RRCKI and Yandex
LLC (Russia), CSCS (Switzerland), IFIN-HH (Romania), CBPF (Brazil),
PL-GRID (Poland) and OSC (USA).
We are indebted to the communities behind the multiple open-source
software packages on which we depend.
Individual groups or members have received support from
AvH Foundation (Germany);
EPLANET, Marie Sk\l{}odowska-Curie Actions and ERC (European Union);
ANR, Labex P2IO and OCEVU, and R\'{e}gion Auvergne-Rh\^{o}ne-Alpes (France);
Key Research Program of Frontier Sciences of CAS, CAS PIFI, and the Thousand Talents Program (China);
RFBR, RSF and Yandex LLC (Russia);
GVA, XuntaGal and GENCAT (Spain);
the Royal Society
and the Leverhulme Trust (United Kingdom).

\appendix
\section{Time-dependent terms}
\label{app:terms}

In Table~\ref{tab:terms},
$\delta_S$ and $\delta_{SS}$ are the strong phases of the $P\to V\kern-0.1em S$ and $P\to S\kern-0.1em S$ processes, respectively. 
The $P$-wave strong phases are dependent on $\delta_{\parallel}$ and \dz.
%defined to be $\delta_1\equiv\delta_\perp-\delta_\parallel$ and $\delta_2\equiv\delta_\perp-\delta_0$,
%with the notation $\delta_{2,1}\equiv\delta_2-\delta_1=\dpa-\dz$.

\begin{sidewaystable}
\caption{Coefficients of the time-dependent terms and angular functions used in Eq.~\ref{eq:pdf}. Amplitudes are defined at $t=0$.}
\newcolumntype{C}{>{$}c<{$}}
\resizebox{\textwidth}{!}{
%\[
%\tiny
  \begin{tabular}{C|C|C|C|C|C|C}
i     & N_i                           & a_i                & b_i                         & c_i               & d_i                  & f_i \\ \hline
1     & |A_0|^2                       & 1+\lambz^2         & -2\lambz\cos (\phiz)        & 1-\lambz^2        & 2\lambz\sin(\phiz)   & 4\cos^2\theta_1\cos^2\theta_2 \\\hline
2       & |A_\parallel |^2            & 1+\lambpa^2        & -2\lambpa\cos (\phipa)      & 1-\lambpa^2       & 2\lambpa\sin(\phipa) & \sin^2\theta_1\sin^2\theta_2(1{+}\cos2\Phi) \\\hline
3       & |A_\perp |^2                & 1+\lambpe^2        & 2\lambpe\cos (\phipe)       & 1-\lambpe^2       & -2\lambpa\sin(\phipe)& \sin^2\theta_1\sin^2\theta_2(1{-}\cos2\Phi) \\ \hline
    4       & \frac{|A_\parallel||A_\perp |}{2}   & \begin{array}{c}\sin(\dpa-\dpe ) -\lambpa\lambpe\cdot \\\sin(\dpa-\dpe-\phipa+\phipe ) \end{array}
                                      & \begin{array}{c}-\lambpa\sin(\dpa-\dpe -\phipa )\\ +\lambpe\sin(\dpa-\dpe+\phipe ) \end{array}
                                      & \begin{array}{c}\sin(\dpa-\dpe ) +\lambpa\lambpe\cdot\\\sin(\dpa-\dpe-\phipa+\phipe ) \end{array}
				      & \begin{array}{c}\lambpa\cos(\dpa-\dpe-\phipa )\\+\lambpe\cos(\dpa-\dpe+\phipe ) \end{array}
				      & -2\sin^2\theta_1\sin^2\theta_2\sin 2\Phi \\\hline
    5       & \frac{|A_\parallel||A_0|}{2}        & \begin{array}{c}\cos(\dpa-\dz ) +\lambpa\lambz\cdot \\\cos(\dpa-\dz-\phipa+\phiz ) \end{array}
                                      & \begin{array}{c}-\lambpa\cos(\dpa-\dz -\phipa )\\ +\lambz\cos(\dpa-\dz+\phiz ) \end{array}
                                      & \begin{array}{c}\cos(\dpa-\dz ) -\lambpa\lambz\cdot\\\sin(\dpa-\dz-\phipa+\phiz ) \end{array}
				      & \begin{array}{c}-\lambpa\sin(\dpa-\dz-\phipa )\\+\lambz\sin(\dpa-\dz+\phiz ) \end{array}
				      & \sqrt{2}\sin2\theta_1\sin2\theta_2\cos\Phi \\\hline
    6       & \frac{|A_0||A_\perp |}{2}           & \begin{array}{c}\sin(\dz-\dpe ) -\lambz\lambpe\cdot \\\sin(\dz-\dpe-\phiz+\phipe ) \end{array}
                                      & \begin{array}{c}-\lambz\sin(\dz-\dpe -\phiz )\\ +\lambpe\sin(\dz-\dpe+\phipe ) \end{array}
                                      & \begin{array}{c}\sin(\dz-\dpe ) +\lambz\lambpe\cdot\\\sin(\dz-\dpe-\phiz+\phipe ) \end{array}
				      & \begin{array}{c}\lambz\cos(\dz-\dpe-\phiz )\\+\lambpe\cos(\dz-\dpe+\phipe ) \end{array}
				      & -\sqrt{2}\sin2\theta_1\sin2\theta_2\sin\Phi\\\hline
7       & |A_{SS}|^2                  & 1+\lambss^2         & -2\lambss\cos (\phissw)        & 1-\lambss^2       & 2\lambss\sin(\phissw)& \frac{4}{9} \\\hline
8       & |A_{S}|^2                   & 1+\lambs^2          & 2\lambs\cos (\phisw)           & 1-\lambs^2        & -2\lambs\sin(\phisw) & \frac{4}{3}(\cos\theta_1+\cos\theta_2)^2 \\\hline
    9       & \frac{|A_S||A_{SS}|}{2}             & \begin{array}{c}\cos(\ds-\dss ) -\lambs\lambss\cdot \\\cos(\ds-\dss-\phisw+\phissw ) \end{array}
                                      & \begin{array}{c}\lambs\cos(\ds-\dss -\phisw )\\ +\lambss\cos(\ds-\dss+\phissw ) \end{array}
                                      & \begin{array}{c}\cos(\ds-\dss ) +\lambs\lambss\cdot\\\sin(\ds-\dss-\phisw+\phissw ) \end{array}
				      & \begin{array}{c}\lambs\sin(\ds-\dss-\phisw )\\+\lambss\sin(\ds-\dss+\phissw ) \end{array}
				      & \frac{8}{3\sqrt{3}}(\cos\theta_1+\cos\theta_2) \\\hline
    10      & \frac{|A_0||A_{SS}|}{2}             & \begin{array}{c}\cos(\dz-\dss ) +\lambz\lambss\cdot \\\cos(\dz-\dss-\phiz+\phissw ) \end{array}
                                      & \begin{array}{c}-\lambz\cos(\dz-\dss -\phiz )\\ +\lambss\cos(\dz-\dss+\phissw ) \end{array}
                                      & \begin{array}{c}\cos(\dz-\dss ) -\lambz\lambss\cdot\\\sin(\dz-\dss-\phiz+\phissw ) \end{array}
				      & \begin{array}{c}-\lambz\sin(\dz-\dss-\phiz )\\+\lambss\sin(\dz-\dss+\phissw ) \end{array}
				      & \frac{8}{3}\cos\theta_1\cos\theta_2\\\hline
    11      & \frac{|A_\parallel||A_{SS}|}{2}     & \begin{array}{c}\cos(\dpa-\dss ) +\lambpa\lambss\cdot \\\cos(\dpa-\dss-\phipa+\phissw ) \end{array}
                                      & \begin{array}{c}-\lambpa\cos(\dpa-\dss -\phipa )\\ +\lambss\cos(\dpa-\dss+\phissw ) \end{array}
                                      & \begin{array}{c}\cos(\dpa-\dss ) -\lambpa\lambss\cdot\\\sin(\dpa-\dss-\phipa+\phissw ) \end{array}
				      & \begin{array}{c}-\lambz\sin(\dpa-\dss-\phipa )\\+\lambss\sin(\dpa-\dss+\phissw ) \end{array}
				      & \frac{4\sqrt{2}}{3}\sin\theta_1\sin\theta_2\cos\Phi\\\hline
    12      & \frac{|A_\perp||A_{SS}|}{2}         & \begin{array}{c}\sin(\dpe-\dss ) -\lambpe\lambss\cdot \\\sin(\dpe-\dss-\phipe+\phissw ) \end{array}
                                      & \begin{array}{c}\lambpe\sin(\dpe-\dss -\phipe )\\ -\lambss\sin(\dpe-\dss+\phissw ) \end{array}
                                      & \begin{array}{c}\sin(\dpe-\dss ) +\lambpe\lambss\cdot\\\sin(\dpe-\dss-\phipe+\phissw ) \end{array}
				      & \begin{array}{c}-\lambpe\cos(\dpe-\dss-\phipe )\\-\lambss\cos(\dpe-\dss+\phissw ) \end{array}
				      & -\frac{4\sqrt{2}}{3}\sin\theta_1\sin\theta_2\sin\Phi\\\hline
    13      & \frac{|A_0||A_{S}|}{2}              & \begin{array}{c}\cos(\dz-\ds ) -\lambz\lambs\cdot \\\cos(\dz-\ds-\phiz+\phisw ) \end{array}
                                      & \begin{array}{c}-\lambz\cos(\dz-\ds -\phiz )\\ -\lambs\cos(\dz-\ds+\phisw ) \end{array}
                                      & \begin{array}{c}\cos(\dz-\ds ) +\lambz\lambs\cdot\\\sin(\dz-\ds-\phiz+\phisw ) \end{array}
				      & \begin{array}{c}-\lambz\sin(\dz-\ds-\phiz )\\-\lambs\sin(\dz-\dss+\phisw ) \end{array}
				      & \begin{array}{c}\frac{8}{\sqrt{3}} \cos\theta_1\cos\theta_2\\\times (\cos\theta_1 + \cos\theta_2)\end{array}\\\hline
    14      & \frac{|A_\parallel||A_{S}|}{2}      & \begin{array}{c}\cos(\dpa-\ds ) -\lambpa\lambs\cdot \\\cos(\dpa-\ds-\phipa+\phisw ) \end{array}
                                      & \begin{array}{c}-\lambpa\cos(\dpa-\ds -\phipa )\\ -\lambs\cos(\dpa-\ds+\phisw ) \end{array}
                                      & \begin{array}{c}\cos(\dpa-\ds ) +\lambpa\lambs\cdot\\\sin(\dpa-\ds-\phipa+\phisw ) \end{array}
				      & \begin{array}{c}-\lambpa\sin(\dpa-\ds-\phipa )\\-\lambs\sin(\dpa-\dss+\phisw ) \end{array}
				      & \begin{array}{c}\frac{4\sqrt{2}}{\sqrt{3}} \sin\theta_1\sin\theta_2\\\times (\cos\theta_1 + \cos\theta_2)\cos\Phi\end{array}\\\hline
    15      & \frac{|A_\perp||A_{S}|}{2}          & \begin{array}{c}\sin(\dpe-\ds ) +\lambpe\lambs\cdot \\\sin(\dpe-\ds-\phipe+\phisw ) \end{array}
                                      & \begin{array}{c}\lambpe\sin(\dpe-\ds -\phipe )\\ +\lambs\sin(\dpe-\ds+\phisw ) \end{array}
                                      & \begin{array}{c}\sin(\dpe-\ds ) -\lambpe\lambs\cdot\\\sin(\dpe-\ds-\phipe+\phisw ) \end{array}
				      & \begin{array}{c}-\lambpe\cos(\dpe-\ds-\phipe )\\+\lambs\cos(\dpe-\ds+\phisw ) \end{array}
				      & \begin{array}{c}-\frac{4\sqrt{2}}{3}\sin\theta_1\sin\theta_2\\\times (\cos\theta_1 + \cos\theta_2)\sin\Phi\end{array}\\%\hline
\end{tabular}
%\]
}

\label{tab:terms}
\end{sidewaystable}

\clearpage
\section{Correlation matrices for the time-dependent fits}
\label{app:corr}

\begin{table}[h]
  \caption{\small \label{tab:corr} Statistical correlation matrix of the time-dependent fit.
  }
\begin{center}
\begin{tabular}{c|cccd{2}d{2}d{2}}
 & $\dpa$ & $\ape$ & $\dpe$ & \multicolumn{1}{c}{$\az$} & \multicolumn{1}{c}{$|\lambda|$} &\multicolumn{1}{c}{\phisPP} \\ \hline
 \dpa& 1.00& 0.14& 0.13& -0.03& 0.02& 0.01 \\
 $\ape$ & & 1.00& 0.01& -0.45& 0.00& -0.03 \\
 $\dpe$ &  & & 1.00& 0.00& -0.26& -0.15 \\
 $\az$ &  &  & & 1.00& -0.01& 0.01 \\
 $|\lambda|$ &  &  &  & & 1.00& -0.05 \\
 $\phisPP$ &  &  &  &  & & 1.00 \\
\end{tabular}
\end{center}
\end{table}

\begin{table}[h]
 \caption{\small Statistical correlation matrix of the time-dependent fit 
  in which \CP violation is polarisation dependent.}
\begin{center}
\begin{tabular}{c|cccd{2}d{2}d{2}}
 & $\dpa$ & $\ape$ & $\dpe$ & \multicolumn{1}{c}{$\az$} & \multicolumn{1}{c}{\phipa} &\multicolumn{1}{c}{\phipe} \\ \hline
 \dpa& 1.00& 0.13& 0.13& -0.02& 0.58& 0.41 \\
 $\ape$ & & 1.00& 0.03& -0.45& 0.00& 0.01 \\
 $\dpe$ &  & & 1.00& 0.00& 0.08& 0.13 \\
 $\az$ &  &  & & 1.00& 0.00& 0.01 \\
 $|\lambda|$ &  &  &  & & 1.00& 0.71 \\
 $\phisPP$ &  &  &  &  & & 1.00 \\
\end{tabular}
\end{center}
%\begin{center}
%\begin{tabular}{c|cccccccccccc}
% & $\dpa$ & $\ape$ & $\dpe$ & $\az$ & $\phipa$ & $\phipe$\\ \hline
  %\dpa& 1.00& 0.13& 0.13& -0.02& \bf{ 0.58 } & 0.41 \\
%  \dpa& 1.00& 0.13& 0.13& $-$0.02&  0.58  & 0.41 \\
%\ape & & 1.00& 0.03& $-$0.45& 0.00& 0.01 \\
%\dpe &  & & 1.00& $-$0.00& 0.08& 0.13 \\
%\az &  &  & & 1.00& $-$0.00& $-$0.00 \\
%\phipa &  &  &  & & 1.00& \bf{ 0.71 }  \\
%\phipa &  &  &  & & 1.00&  0.71  \\
%\phipe &  &  &  &  & & 1.00 \\
%\end{tabular}
% \end{center}
 \label{tab:corrDep}
 \end{table}

\clearpage
\addcontentsline{toc}{section}{References}
%\setboolean{inbibliography}{true}
\bibliographystyle{LHCb}
\bibliography{main,standard,refs,LHCb-PAPER,LHCb-CONF,LHCb-DP,LHCb-TDR}

\newpage
% LHCb collaboration author list
% Data extracted on July 8th, 2019 at 9:03pm for reference date 14-Mar-2019
\centerline
{\large\bf LHCb collaboration}
\begin
{flushleft}
\small
R.~Aaij$^{30}$,
C.~Abell{\'a}n~Beteta$^{47}$,
B.~Adeva$^{44}$,
M.~Adinolfi$^{51}$,
C.A.~Aidala$^{78}$,
Z.~Ajaltouni$^{8}$,
S.~Akar$^{62}$,
P.~Albicocco$^{21}$,
J.~Albrecht$^{13}$,
F.~Alessio$^{45}$
% ...

{\footnotesize \it

% ...
$ ^{8}$Universit{\'e} Clermont Auvergne, CNRS/IN2P3, LPC, Clermont-Ferrand, France\\
% ...
$ ^{13}$Fakult{\"a}t Physik, Technische Universit{\"a}t Dortmund, Dortmund, Germany\\
% ...
$ ^{21}$INFN Laboratori Nazionali di Frascati, Frascati, Italy\\
% ...
$ ^{30}$Nikhef National Institute for Subatomic Physics, Amsterdam, Netherlands\\
% ...
$ ^{44}$Instituto Galego de F{\'\i}sica de Altas Enerx{\'\i}as (IGFAE), Universidade de Santiago de Compostela, Santiago de Compostela, Spain\\
$ ^{45}$European Organization for Nuclear Research (CERN), Geneva, Switzerland\\
% ...
$ ^{47}$Physik-Institut, Universit{\"a}t Z{\"u}rich, Z{\"u}rich, Switzerland\\
% ...
$ ^{51}$H.H. Wills Physics Laboratory, University of Bristol, Bristol, United Kingdom\\
% ...
$ ^{62}$University of Cincinnati, Cincinnati, OH, United States\\
% ...
$ ^{65}$Syracuse University, Syracuse, NY, United States\\
% ...
$ ^{78}$University of Michigan, Ann Arbor, United States, associated to $^{65}$\\
\bigskip
% ...
}
\end{flushleft}

\end{document}